%% file: main.tex
\documentclass[twocolumn]{aastex63}
\usepackage{amsmath}
\usepackage{csvsimple}
\usepackage{xcolor}
\usepackage{stix}
\DeclareUnicodeCharacter{0360}{\eth}

\shorttitle{Solar RV Comparison}
\shortauthors{Zhao, Dumusque, Ford, Llama, Mortier, et al.}

\newcommand{\project}[1]{\textsl{#1}}
\newcommand{\acronym}[1]{{\small{#1}}}
\newcommand{\code}[1]{\texttt{#1}}
\newcommand{\expres}{\acronym{EXPRES}}
\newcommand{\lost}{\acronym{LOST}}

\newcommand{\ldt}{\acronym{LDT}}
\newcommand{\tng}{\acronym{TNG}}

\newcommand{\espresso}{\acronym{ESPRESSO}}
\newcommand{\neid}{\acronym{NEID}}
\newcommand{\harps}{\acronym{HARPS}}
\newcommand{\helios}{\acronym{HELIOS}}
\newcommand{\harpsn}{\acronym{HARPS-N}}
\newcommand{\lcst}{\acronym{LCST}}
\newcommand{\maroonx}{\acronym{MAROON-X}}
\newcommand{\kpf}{\acronym{KPF}}
\newcommand{\pepsi}{\acronym{PEPSI}}
\newcommand{\gianob}{\acronym{GIANO-B}}

\newcommand{\poet}{\acronym{PoET}}
\newcommand{\aboras}{\acronym{ABORAS}}
\newcommand{\nirps}{\acronym{NIRPS}}
\newcommand{\bison}{\acronym{BiSON}}
\newcommand{\gong}{\acronym{GONG}}

\newcommand{\selenite}{\code{SELENITE}}
\newcommand{\yarara}{\project{\acronym{YARARA}}}
\newcommand{\scalpels}{\project{\acronym{SCALPELS}}}

\newcommand{\cms}{\mbox{cm s\textsuperscript{-1}}}
\newcommand{\ms}{\mbox{m s\textsuperscript{-1}}}

\newcommand{\vmax}{$\nu_\mathrm{max}$}

\newcommand{\harpsncolor}{red}

\newcommand{\exprescolor}{yellow}
\newcommand{\neidcolor}{orange}

\begin{document}

\title{The Extreme Stellar-Signals Project III. Combining Solar Data from \harps, \harpsn, \expres, and \neid}
\input{authors}

\begin{abstract}
We present an analysis of Sun-as-a-star observations from four different high-resolution, stabilized spectrographs---\harps, \harpsn, \expres, and \neid.  With simultaneous observations of the Sun from four different instruments, we are able to gain insight into the radial velocity precision and accuracy delivered by each of these instruments and isolate instrumental systematics that differ from true astrophysical signals.  With solar observations, we can completely characterize the expected Doppler shift contributed by orbiting Solar System bodies and remove them.  This results in a data set with measured velocity variations that purely trace flows on the solar surface.  Direct comparisons of the radial velocities measured by each instrument show remarkable agreement with residual intra-day scatter of only 15-30~\cms.  This shows that current ultra-stabilized instruments have broken through to a new level of measurement precision that reveals stellar variability with high fidelity and detail.  We end by discussing how radial velocities from different instruments can be combined to provide powerful leverage for testing techniques to mitigate stellar signals.
\end{abstract}


\section{Introduction}

The radial velocity (RV) method of discovering exoplanets measures the center-of-mass motion of a planet's host star as both the planet and star orbit their common center of mass.  This center-of-mass motion induces a periodic Doppler shift on the star's spectrum over time.  Highly-stabilized spectrographs are able to measure this Doppler shift down to sub-meter-per-second levels \citep{pepe2004, cosentino2012, pepe2013, schwab2016, jurgenson2016}.

Changes on a star's surface will introduce spectral variations that can be mistakenly measured as a true center-of-mass shift of the spectra \citep{meunier2021}.  These surface variations include, but are not limited to, p-mode oscillations \citep{mayor2003, bouchy2005, kjeldsen2005, arentoft2008, chaplin2019}, granulation \citep{dravins1982, kjeldsen1995, lindegren2003, dumusque2011-01, meunier2015, cegla2018, lanza2019-03}, and supergranulation \citep{rieutord2010, rincon2018, meunier2019}.  Activity features---such as spots, faculae, and plages \citep{saar1997, hatzes2002, saar2003, desort2007, huelamo2008, boisse2011,  dumusque2011-03, lovis2011, jeffers2013, santos2014, cabot2021, roettenbacher2021}---can also introduce periodic spectral variations as they rotate in and out of view as well as give rise to non-periodic signals as these activity regions and their properties evolve with time.

Measured RV variations due to stellar surface variations contribute scatter or potentially periodic offsets to RV  time series \citep{crass2021}.  Different types of variations will manifest in the spectra in different ways, in different lines, and on different time scales.  Spectral lines affected by these surface variations may be shallower/deeper, appear shifted, or acquire asymmetric wings.  Some spectral lines are likely to be more affected than others depending on the line's wavelength, species, and formation temperature \citep{dumusque2018, wise2018, ning2019, cretignier2020-01, almoulla2022}.  These variations will introduce (potentially periodic) RV scatter of an amplitude far greater than the $\sim$10 \cms\ expected signal for an Earth-like planet orbiting a Sun-like star.

Many methods have been developed to mitigate the variations that arise in RV measurements due to stellar signals \citep[e.g.][]{aigrain2012, jones2017, lafarga2020, rajpaul2020, zhao2020, gilbertson2020-02, collier2021, cretignier2021}.  The previous paper in this series \citep[The \expres\ Stellar Signals Project II\footnote{Note, this is the previous name of this series.  With this installment, the series name is changed to ``The Extreme Stellar Signals Project'' as the analysis expands beyond \expres\ data.},][]{zhao2022} performed a head-to-head comparison of different mitigation methods tested on the same set of real observations for four different stars.  This direct comparison produced a confusing outcome---methods often disagreed on the signals that should be attributed to stellar surface variations, even though all methods used identical data sets.

That exercise highlighted the difficulty of rigorously establishing method performance.  The choice to test methodologies using real vs.\ simulated data sets highlights different strengths and weaknesses of the methods and introduces very different working hypotheses.  Simulated data sets come with the benefit of knowing the ``answer at the back of the book''---in other words, the resulting data set is one for which the structure of the injected signals are known. It is also known whether there are any true center-of-mass shifts (simulated planetary signals) that mitigation methods need to preserve.  As with all simulations, however, it is unclear whether the simulated signals are physically motivated and complete enough to be representative of a real-case scenario.  Using real data sets, however, introduces non-astrophysical sources of systematics (e.g., instrumental variation, different observing conditions/limitations, etc.) and the potential of un-detected planetary signals, which can lead to confusion and misinterpretations of results.

Sun-as-a-star observations---i.e., disk-integrated observations of the Sun as if it were a star---have many of the benefits of both simulated data and real observations.  Because they are observations of a real star, solar observations are sure to capture realistic manifestations of stellar variability.  True center-of-mass shifts are well-understood for the Sun, as Solar System bodies are characterized well beyond the limits of current RV capabilities.  Observing the Sun is also relatively inexpensive as observations occur during the day when the spectrographs are not otherwise used.  With such a bright source, high signal-to-noise ratio (SNR) observations can easily be achieved as well as temporal sampling on the timescale of minutes.

Several planet-hunting spectrographs therefore have solar feeds---i.e.\ the High-Accuracy Radial-velocity Planet Searcher (\harps), the High-Accuracy Radial-velocity Planet Searcher in the North \citep[\harpsn;][]{phillips2016}, the EXtreme PREcision SPectrograph \citep[\expres;][]{llama2023}, the NN-explore Exoplanet Investigations with Doppler spectroscopy instrument \citep[\neid;][]{lin2022}, and the Keck Planet Finder \citep[\kpf;][]{rubenzahl2023}.  Other instruments have plans to add a solar feed---e.g.\ on the \'Echelle Spectrograph for Rocky Exoplanet and Stable Spectroscopic Observations \citep[\espresso;][]{leite2022} or the M dwarf Advanced Radial velocity Observer Of Neighboring eXoplanets (\maroonx).

Solar observations also feature some unique disadvantages when compared with nighttime stellar observations.  As a resolved source, solar observations must contend with differences in observing conditions (e.g., cloud coverage, airmass, etc.) across the disk of the Sun.  The Earth's relative velocity with respect to the Sun is lower than the relative velocity of the Earth to other stars, meaning telluric lines, from absorption by Earth's atmosphere, do not shift as much relative to solar lines as stellar lines.  Observations that span a variety of stellar types and stellar parameters beyond the Sun will ultimately be necessary to more holistically understand how surface variations present in stabilized spectral data.  However, solar data is a valuable starting point for understanding stellar signals as measured by precise radial-velocity (PRV) spectrographs.

We combine data from four different PRV spectrographs---\harps\ \citep{pepe2004}, \harpsn\ \citep{cosentino2012}, \expres\ \citep{jurgenson2016}, and \neid\ \citep{schwab2016}.  Having concurrent data of the same source allows us to benchmark each instrument against the other three.  This can reveal how differences in instrument design, instrument systematics, and data reduction processes might affect how the same stellar signals manifest in the data collected by each instrument.

Here we present and discuss this combined solar data set.  In Section~\ref{sec:data}, we describe the different instruments, their corresponding data sets, and quantify the degree to which they overlap with one another.  We also discuss how each instrument's pipeline contends with some of the disadvantages of observing the Sun.  Section~\ref{sec:methods} explains how the RVs from the different instruments are compared across different timescales; results from this comparison are outlined in Section~\ref{sec:results}.  We end with conclusions and a discussion of future directions in Section~\ref{sec:discussion}.

\section{Data}\label{sec:data}

We combine data from \harps, \harpsn, \expres, and \neid.  We use observations collected between May 25 and June 23, 2021, which were selected as dates that contained a large number of overlapping data from the four instruments.  Only exposures taken at an airmass less than 2.2 are used.  Including observations taken at an airmass greater than 2.2 introduces significant error from atmospheric extinction.  With the Sun, higher arimasses also increases the differential extinction effect of the resolved disk of the Sun.  Enacting an airmass cut that is harsher, however, significantly reduce the amount of overlapping data we have between different instruments.

\begin{figure}[tp]
\centering
\includegraphics[width=.45\textwidth]{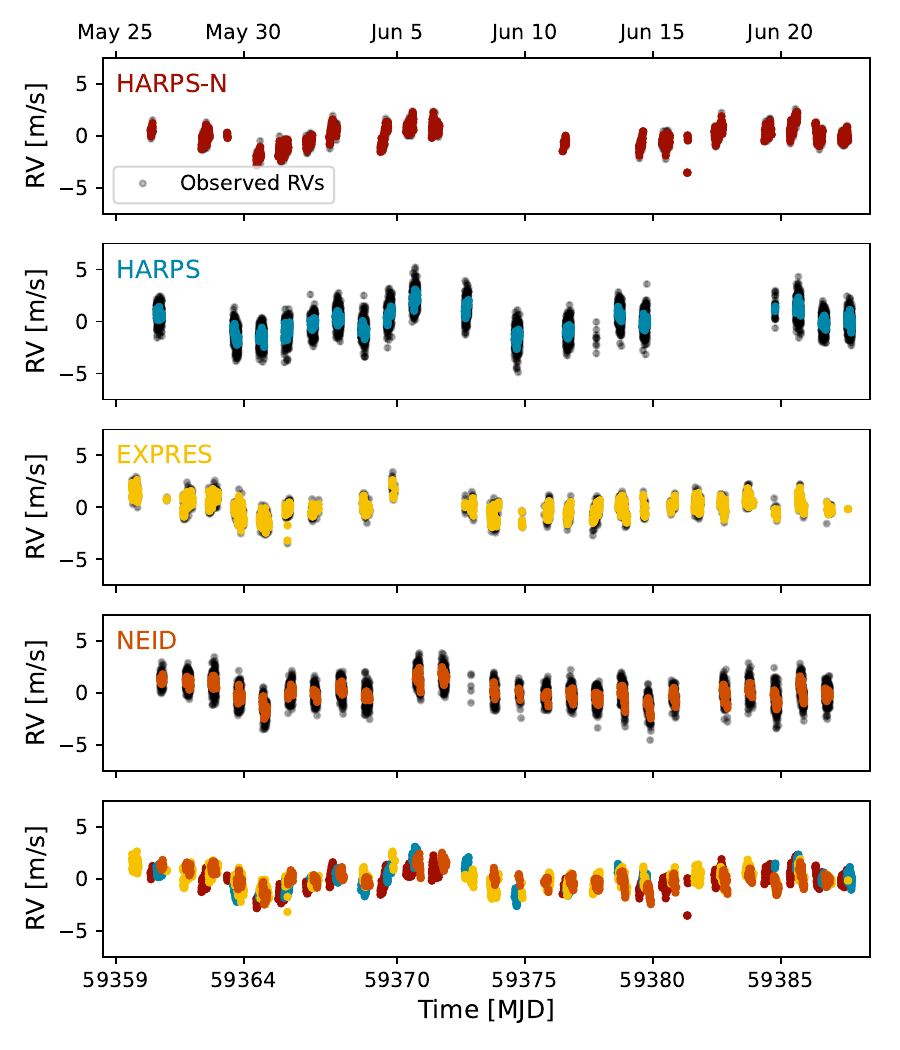}
\caption{Solar RVs from \harpsn, \harps, \expres, and \neid.  Observed RVs are shown as black points; colored points represent the RVs binned to shared timestamps with an interval of five minutes throughout a day.  The color assigned to the data for each instrument is consistent throughout this work.  The bottom most subplot overlays the binned RVs from all instruments (with a separate offset applied to each instrument).  All instruments trace similar long-term RV variations.}
\label{fig:rvs}
\end{figure}

The observations from each instrument are plotted in Figure~\ref{fig:rvs} as black points.  Colored points show the observed RVs from each instrument binned to five-minute intervals (see Section~\ref{sec:binning} for a detailed description of the binning process).  We will only use ``observations'' to refer to the RV measurements returned by each instrument's data reduction pipeline (i.e., the binned RVs will be referred to as ``binned RVs'' or ``binned points'' rather than observations).

This analysis makes use of the RVs from each instrument's default data reduction pipeline (DRP).  In other words, the RVs used are produced by the same pipeline that is used to extract the nightly data for facility users of each instrument with some additional Sun-specific corrections (e.g., implementing barycentric corrections as relevant to solar observations, accounting for the Sun being a resolved source, and others discussed further below).  The RVs used here have had known Solar System planet signals removed using the known ephemerides of these objects as curated by the Jet Propulsion Laboratory (JPL) \citep{park2021}.  These shifts were removed at the wavelength level for each spectrum.  For more details about each instrument's DRP, see \cite{dumusque2021} for \harps\ and \harpsn, \cite{petersburg2020} for \expres, and \url{https://neid.ipac.caltech.edu/docs/NEID-DRP/} for \neid.

\begin{figure}[tp]
\centering
\includegraphics[width=.45\textwidth]{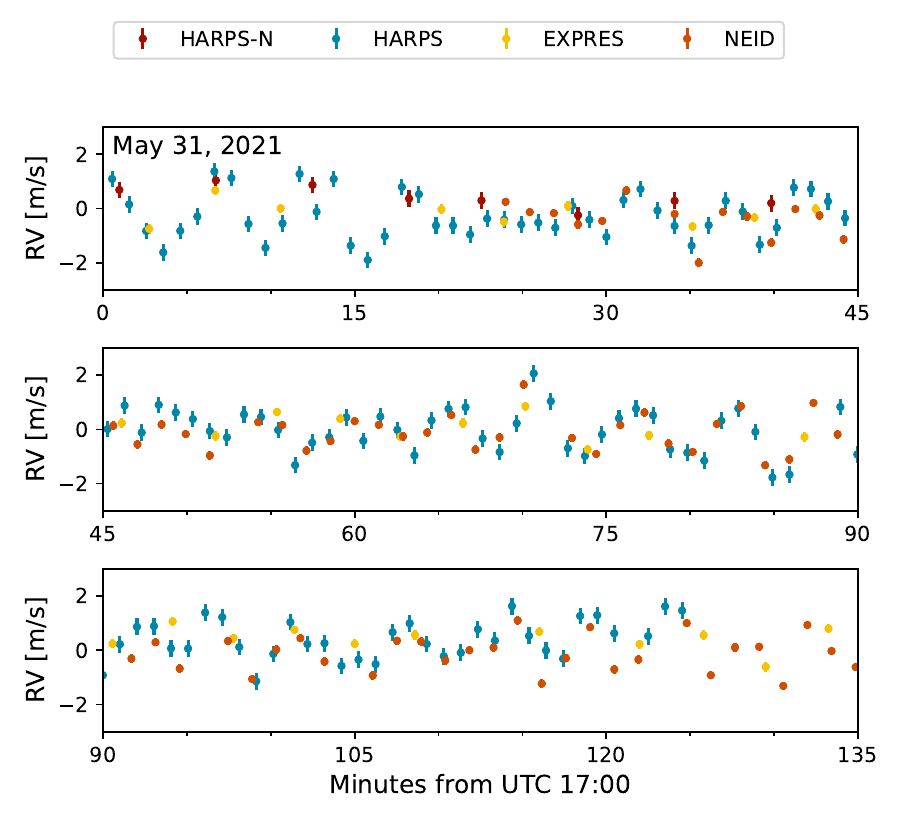}
\caption{Observed RVs from \harpsn, \harps, \expres, and \neid.  DRP RVs (i.e., not binned RVs) are shown over two hours on May 31, 2021 from UTC 17:00 to UTC 19:15.  A separate RV offset, calculated as the median of all RVs from that instrument taken on the day, has been applied to each instrument.  The x-axis is given in minutes from UTC 17:00 with minor tick marks every 5-minutes (which is close to the \vmax\ for the Sun).  All four instruments trace similar short-term RV variations.}
\label{fig:og_rvs}
\end{figure}

Figure~\ref{fig:og_rvs} shows a two-hour subset of the observed RVs taken on May 31, 2021 from each instrument (i.e., the figure shows the observed RVs as provided by each instrument's pipeline, not the RVs binned to shared timestamps).  As it is impractical to show all the data at this high time resolution, we show here the period of time with the greatest amount of overlap between the four instruments.  This is also the only time span within the month for which there are good quality solar observations from all four instruments (from $\sim$17.4 to 17.7 UTC time of day).  Figure~\ref{fig:og_rvs}, therefore, shows the optimal time frame for a direct comparison of the DRP RVs from all four instruments on short-time scales.  Comparisons that take into account all times when instruments overlap across the entire month are done using the binned RVs and are discussed more below.

Table~\ref{tab:data} gives instrument and data properties for each data set.  For each instrument, we list the start of science operations (i.e., the time at which each instrument is considered to have been fully commissioned), the start of regular observations for each instrument's corresponding solar telescope, the full wavelength range of instrument, and the median resolution across each instrument's detector.  We also compare a subset of hardware components across solar telescopes that differ between instruments.  All four solar telescopes employ a 75-mm achromatic doublet lens from Edmumd Optics (\neid\ has a slightly different coating to address its redder wavelength range), and the light is combined via a 2'' Polytetrafluoroethylene (PTFE) integrating sphere from Thorlabs.  \expres\ is the only telescope of the four to use an equatorial mount.  \harps, \harpsn, and \expres\ all use the same type of telescope dome (\neid's solar telescope has no dome).

To capture the number and cadence of observations, Table~\ref{tab:data} gives the amount of data, exposure time, and average readout time for each instrument.  For each data set, we give both the total number of observations and the total number of binned data points.  The number of days (out of the 29 included in this analysis) on which there is data from each instrument is also given.  Note, \expres\ terminates exposures at a set SNR rather than at a fixed time; therefore an average exposure time is cited for \expres\ in Table~\ref{tab:data}.

\begin{table*}[tb]
\scriptsize
\caption{Instrument/Data Properties}
\label{tab:data}
\begin{center}
\begin{tabular}{l c c c c}
\hline
\hline
Parameter & \harps & \harpsn & \expres & \neid \\
\hline
\textbf{Instrument} &  &  &  &  \\
Solar Telescope Name & \helios & \lcst & \lost & \neid\ Solar Feed \\
Location & La Silla, Chile & La Palma, Spain & Flagstaff, AZ, USA & Kitt Peak, AZ, USA \\
\hspace{1.5em}Latitude & 29$^{\circ}$15'27'' S & 28$^{\circ}$45'49'' N & 34$^{\circ}$44'40'' N & 31$^{\circ}$57'30'' N \\
\hspace{1.5em}Longitude & 70$^{\circ}$44'15'' W & 17$^{\circ}$53'41'' W & 111$^{\circ}$25'19'' W & 111$^{\circ}$35'48'' W \\
Start of Spectrograph Science Operations & 2003 & 2012 & 2019 & 2021 \\
Start of Solar Observations & 2018 & 2015 & 2020 & 2020 \\
Full Wavelength Range [nm] & $378-691$ & $383-690$ & $390-780$ & $380-1046$ \\
Median Resolution & 115,000 & 118,000 & 137,000 & 120,000 \\
\emph{Solar Telescope Components} &  &  &  & \\ 
Aperture of Solar Telescope Lens [mm] & 75 & 75 & 75 & 75 \\
Lens Coating & MgF\textsubscript{2} & MgF\textsubscript{2} & MgF\textsubscript{2} & VIS-NIR\textsuperscript{(a)} \\
Solar Telescope Mount & Alt/Az & Alt/Az & Equatorial & Alt/Az \\
Dome Material & Polymethyl Methacrylate & Polymethyl Methacrylate & Polymethyl Methacrylate & None \\
Fiber Feed Length [m] & $\sim$30 & $\sim$20 & $\sim$80 & $\sim$45 \\
\hline
\textbf{Data} &  &  &  & \\ 
\emph{Amount of Data\textsuperscript{(b)}} &  &  &  & \\ 
Total Observations & 4570 & 1082 & 1459 & 3617 \\ 
Binned RVs & 849 & 1176 & 1218 & 1059 \\ 
Days w/ Data (Out of 29) & 19 & 19 & 26 & 25 \\ 
Avg. Obs./Day & 157 & 37 & 50 & 124 \\ 
\emph{Cadence} &  &  &  & \\ 
Exposure Time [s] & 30 & 300 & 178\textsuperscript{(c)} & 55 \\Dead Time [s] &  & 26 & 52 & 28 \\ 
\emph{Data Quality} &  &  &  & \\ 
Average RV Error [m/s] & 0.3 & 0.29 & 0.19 & 0.16 \\ 
Average Binned RV Error [m/s] & 0.09 & 0.21 & 0.12 & 0.06 \\ 
SNR at 4200 \AA & 140 & 130 & 100 & 250 \\ 
SNR at 4750 \AA & 280 & 220 & 270 & 420 \\ 
SNR at 5500 \AA & 400 & 290 & 480 & 570 \\ 
SNR at 6250 \AA & 420 & 310 & 620 & 630 \\ 
SNR at 6650 \AA & 410 & 300 & 660 & 640 \\ 
\emph{Start/Stop Criteria \textsuperscript{(d)}} &  &  &  & \\
Start Airmass & 5.1 & 2.9 & 2.7 & \emph{$\sim$1.3} \\
Stop Airmass & 4.6 & \emph{$\sim$1.3} & \emph{$\sim$2.0} & \emph{$\sim$1.3} \\
Start Time [UTC] & \emph{$\sim$12:37} & \emph{$\sim$7:55} & \emph{$\sim$13:11} & 16:31 \\
Stop Time [UTC] & \emph{$\sim$20:40} & 16:00 & (+1) 00:00 & 22:30 \\
\hline
\end{tabular}
\end{center}
{\raggedright (a) The \neid\ lens makes use of the proprietary VIS-NIR anti-reflection coating from Edmund Optics \citep{lin2022} \\
(b) Specifically the amount of data used in this analysis, which only includes data collected between May 25-June 23, 2021 \\
(c) An average exposure time is given here because \expres\ uses a set SNR rather than a fixed exposure time \citep{petersburg2018} \\ (d) Italicized values indicate that those values were calculated as opposed to the values actually used to start/stop observations \par}
\end{table*}

\begin{figure*}[tb]
\centering
\includegraphics[width=.75\textwidth]{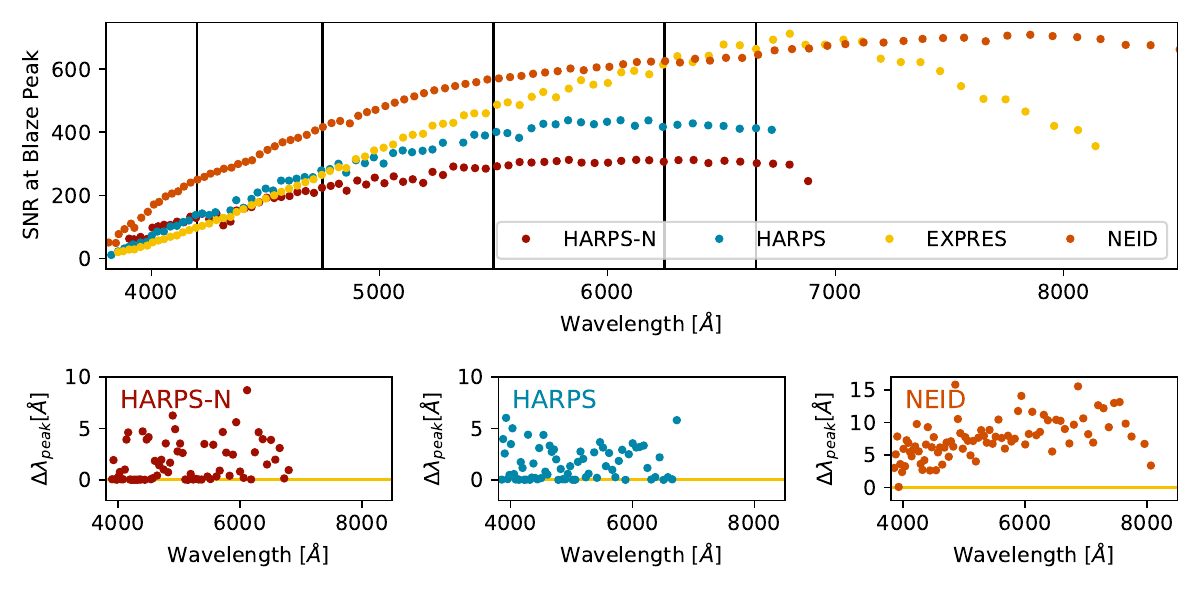}
\caption{\textbf{Top row}: Per-exposure solar SNR at the blaze peak of every \'echelle order for a representative observation from each instrument.  Black vertical lines mark the wavelengths for which the average SNR is reported in Table~\ref{tab:data}.  Note, \neid's wavelength range extends further red beyond the edge of the plot.  \textbf{Bottom row}: Difference between the wavelength of an instrument's blaze peak, $\lambda_\mathrm{peak}$, for each order.  From left to right, \harpsn, \harps, and \neid\ are compared to the blaze peak wavelength of \expres.  For \harpsn\ and \harps, the $\lambda_\mathrm{peak}$ of each order falls on average within 2 \AA\ of the \expres\ $\lambda_\mathrm{peak}$; between \neid\ and \expres\ the $\lambda_\mathrm{peak}$ is on average within 7 \AA.}
\label{fig:throughput}
\end{figure*}

\begin{figure*}[tp]
\centering
\includegraphics[width=.8\textwidth]{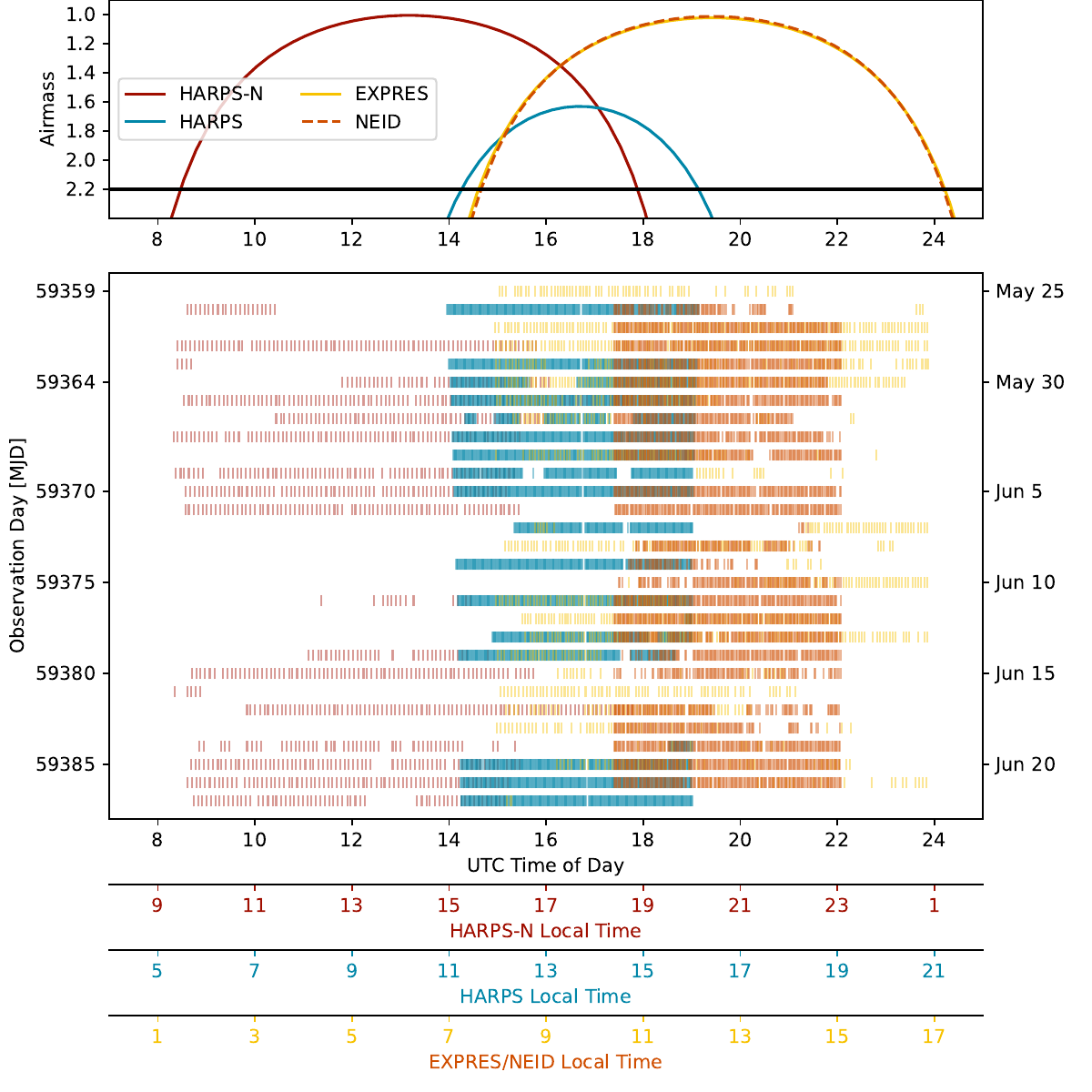}
\caption{The timestamps of observations from each instrument as a function of time of day.  \textbf{Top}: Airmass of the Sun in June as a function of UTC time of day as seen from each of the four instruments.  The \expres\ and \neid\ curves largely overlap.  The analysis presented here uses only data taken at an airmass lower than 2.2 (marked by the horizontal black line).  \textbf{Bottom}: Timestamps of observations from each instrument for each day (y-axis, increasing top to bottom) and UTC time of day (x-axis).  This layout highlights the cadence of each time series as well as the times at which observations from different instruments overlap or when there are gaps in the observations.  To help interpret UTC time of day, the local time at each instrument's location is given on separate axes below this plot.  (Because \expres\ and \neid\ are located in the same time zone, they share a local time axis that is therefore colored \exprescolor\ and \neidcolor.)}
\label{fig:airmass}
\end{figure*}

The average quality of observations from each instrument is given in terms of the average analytic RV error and the SNR (per 1-D extracted pixel) of a representative observation for a range of wavelengths.  The average analytic RV error given in Table~\ref{tab:data} is directly calculated using just the month of shared data.  The quoted value may therefore differ from the approximate analytic errors expected from each instrument's DRP, which is typically calculated using a larger and more varied set of observations. For \harps\ and \harpsn, the SNR per 1-D extracted pixel is calculated from the flux of each pixel divided by the expected photon noise and detector noise added in quadrature.  In the case of high SNR observations, like of the Sun, the SNR is close to the square root of the flux.  For \expres, the SNR is directly calculated as the square root of the photon count.  For \neid, the SNR is calculated from the ratio of the science fiber's flux and the science fiber's analytic variance, which is given by the error associated with flat-relative optimal extraction as described in \cite{zechmeister2014}.  Similar to the other instruments, the analytic variance calculated by \neid's DRP is dominated by photon noise, especially at high SNR.

Figure~\ref{fig:throughput} shows the average peak SNR within each \'echelle order---i.e., the peak of the blaze function within that order---plotted against the wavelength at which the peak occurs, $\lambda_\mathrm{peak}$.  The average SNR at five wavelengths, as marked in Figure~\ref{fig:throughput} by vertical black lines, is listed in Table~\ref{tab:data}.  The bottom row of subplots in Figure~\ref{fig:throughput} shows the difference in $\lambda_\mathrm{peak}$ for \harpsn, \harps, and \neid\ as compared to the $\lambda_\mathrm{peak}$ of \expres\ in each order.  Between the different instruments, the position of the blaze peak differs on average by less than 2~\AA\ between \expres\ and both \harpsn\ and \harps, which corresponds to less than 1.5-3\% of the order width.  The blaze peak for \neid\ differs from \expres\ by less than 7~\AA\ on average, or 7.5\% of the \neid\ order width.  This agreement is expected as all instruments make use of an R4 \'echelle grating with 31.6 lines per millimeter.

Table~\ref{tab:data} ends with the start/stop criteria used by each instrument, i.e.\ when solar observations are scheduled to begin and end for each day.  \neid\ uses time of day as its start/stop criteria, and observations are fully automated in the instrument computer.  \harps\ and \harpsn\ uses the altitude of the Sun to guide the beginning and end of solar observations.  Observations are manually started by the telescope's day operators, and observations proceed until manually stopped for late-afternoon calibrations.  \expres\ is fully automated and uses the altitude of the Sun to start operations and time of day to stop observing.  Instruments using altitude-based stop criteria are often stopped earlier if the instrument is needed to prepare for nighttime observing.  Start/stop times are given in UTC time of day, not in the local time of the telescope location.

We give the observing window criteria used by each instrument in the table as roman (i.e.\ non-italic) text.  We also compute the corresponding value (i.e., the airmass that corresponds to a start/stop time or the start/stop time that corresponds to an airmass) to allow for easier comparison across all instruments.  These calculated values are presented in the table in italicized text.  The conversions are done for a day in the middle of the shared time series (June 8, 2021).

The timing of the solar observations from each instrument is diagrammed in Figure~\ref{fig:airmass}.  The top of Figure~\ref{fig:airmass} shows the airmass of the Sun in June across a day as seen by each instrument.  This is shown as a function of UTC time of day.  The bottom plot of the figure shows the timestamps of each observation over the 29 days of data used in the analysis.  The day as given by its MJD (left y-axis) or calendar day (right y-axis) increases from top to bottom.  Comparing the timestamps of the observations shows when and how often observations from different instruments overlap as well as gaps in the overlap time.  The corresponding local time for each instrument is given in the additional x-axes at the bottom of the plot.

\subsection{\harps}\label{sec:harps}

The High Accuracy Radial Velocity Planet Searcher, \harps, is a stabilized, fiber-fed, optical ($378-691$~nm) \'echelle spectrograph \citep{pepe2004}.  It was fully commissioned at the 3.6-m telescope at La Silla Observatory, Chile in October 2003. \harps\ has a median resolution of $R\sim115,000$ and an instrument calibration precision of 50~\cms. Daily solar observations show a RV RMS that is below 50~\cms\ (where data has been binned over five minutes to average over the p-mode oscillation signal) demonstrating the short-term stability of this instrument\footnote{when using the implementation of the \espresso\ pipeline for \harps\ data (\url{https://www.eso.org/sci/software/pipelines/espresso/espresso-pipe-recipes.html}), as done in this paper for the Sun. Note that this pipeline is still in development.}.

\harps\ has also demonstrated long-term stability close to 1~\ms. \citet{cretignier2021} shows that observations of Tau Ceti over 13 years gives an RV RMS of 1.18~\ms, where the RVs delivered by the official \harps\ pipeline have been daily-binned to mitigate stellar oscillation and granulation signals.  The RV RMS of the same data set can be reduced to 1.02~\ms\ with the use of post-processing techniques, such as \yarara, to mitigate instrumental systematics that are challenging to correct for at the extraction level \citep{cretignier2021}. \citet{cretignier2023} shows that an RV RMS of smaller than 1~\ms\ can be achieved for $\sim$10~years of observations of four other \harps\ RV standards, and that new planetary candidates around these stars with amplitudes as small as 0.5~\ms\ for periods as long as 600 days can be detected.


The \harps\ Experiment for Light Integrated Over the Sun (\helios) is a 75-mm telescope with a fiber-feed into \harps. It has been observing the Sun as a star since first light in 2018. \helios\ starts observing the Sun when it rises above an altitude of 10$^{\circ}$ and is programmed to stop when the Sun sets below 10$^{\circ}$.  In practice, solar observations are typically stopped much earlier to prepare for nighttime observations.  Exposure times are fixed at 30~s, for which \harps\ reaches an analytical error per observation of 30~\cms.

The pipeline for the \helios\ data is similar to the pipeline for \harpsn, as described in \citet{dumusque2021}. It is largely the same pipeline as used for the nightly stellar data observed with either \espresso\ or \harpsn. \helios\ observes through a plexiglass dome and thus observes even when it is cloudy (although observations are stopped when the guiding camera cannot find the solar disc).  Data quality is therefore assessed a posteriori via a Bayesian mixture-model approach, which takes in the SNR and airmass of the data for a given day, to vet observations during which the Sun was partially obscured by clouds and/or other weather conditions \citep{collier2019}.

This procedure also produces a daily atmospheric extinction coefficient.  This value is subsequently used to correct the solar RVs for differential atmospheric extinction effects that arise because the Sun is a resolved disc in the sky.  Light from the lowest-point of the Sun therefore travels through more of the atmosphere and has a higher airmass than light from  the top of the Sun; this creates a daily downwards trend in the RVs.  Details of this extinction correction and the Bayesian mixture-model as applied to \harps\ and \harpsn\ are described further in \citet{almoulla2023} and \citet{collier2019}, respectively.

\subsection{\harpsn}\label{sec:harpsn}

The High Accuracy Radial Velocity Planet Searcher in the North, \harpsn, is a stabilized, fiber-fed, optical ($383-690$~nm) \'echelle spectrograph \citep{cosentino2012}.  It was fully commissioned at the 3.6-m Telescopio Nazionale Galileo (\tng) at La Palma, Spain in August 2012. \harpsn\ has a median resolution of $R\sim118,000$ and instrument calibration precision of 49~\cms\ (as measured by the median absolute offset between consecutive wavelength solutions). The on-sky performances are very similar to what is obtained with \harps, with a daily RV rms for solar observation (5-minute integration time) at around 40~\cms.

\harpsn\ exhibits similar long-term stability as \harps.  \citet{john2023} show that RV standard star HD~144579 returns an RV RMS of 1.29~\ms\ over 9~years of observations binned nightly, similar to the precision reached by \harps\ for Tau Ceti (as described above in Section~\ref{sec:harps}.  Moving beyond just RV RMS, \citet{john2023} also demonstrate that for HD~144579, planets with periods between 1 and 2000 days have a mean detectability limit of just 0.62~\ms\ with RVs from the \harpsn\ pipeline, and a mean detectability of 0.54~\ms\ for RVs that have been post-processed with \scalpels\ \citep{collier2021}.

The HARPS-N Low-Cost Solar Telescope \citep[\lcst;][]{dumusque2015, phillips2016} is a 75-mm achromatic lens telescope with a fiber-feed into \harpsn. It has been observing the Sun as a star since first light in July 2015. The \harpsn\ \lcst\ starts observing the Sun when it rises above an altitude of 20$^{\circ}$.  Calibrations are taken at the end of the previous observing night. There is a hard stop to observations when the sun is below an altitude of 20$^{\circ}$, but in practice observations are typically stopped every day between 14:00 and 16:00 UTC (15:00-17:00 local time) in preparation for nighttime observations. Data are calibrated using simultaneous observations of a Fabry-Perot etalon.  Exposure times are fixed at 300~s (5~minutes) in an effort to bin over RV variations from solar oscillations \citep[e.g.][]{chaplin2019}.  On a clear day, \harpsn\ reaches an approximate analytical error per observation of 25~\cms.

The pipeline for the \harpsn\ \lcst\ data is described in \citet{dumusque2021}.  As with \harps, the same data quality factor and atmospheric extinction coefficient is calculated to ensure good quality data with \harpsn\ \citep{collier2019}.  For \harpsn, an additional quality cut is made using the ratio between the max and mean counts of the exposure meter for each observation.  Observations with steady illumination of the fiber will return a ratio closer to one.  Observations with a ratio greater than $3\sigma$ away from the mean are cut.

\subsection{\expres}\label{sec:expres}

The EXtreme PREcision Spectrograph, \expres, is a stabilized, fiber-fed, optical ($390-780$~nm) \'echelle spectrograph \citep{jurgenson2016}.  It was fully commissioned at the 4.3-m Lowell Discovery Telescope \citep[\ldt;][]{levine2012} near Flagstaff, AZ in January 2019.  \expres\ has a median resolution of $R\sim137,000$ and an instrument calibration precision of 4-7~\cms, as measured by the RMS of the perceived shifts of 0.5-1 hour of consecutive calibration exposures \citep{blackman2020, zhao2022}.  For the quietest stars, \expres\ returns an on-sky precision of $\sim$60~\cms\ \citep{zhao2022}.

The Lowell Observatory Solar Telescope, \lost, is a 75-mm lens telescope with a fiber-feed into \expres\ that has been observing the Sun as a star since first light in late 2020 (Llama et al.\ in prep.).  Calibration images for reducing solar observations are initiated when the Sun rises above 10 degrees.  Calibration images take about one hour to complete and solar observations start immediately afterwards; at this point, the Sun tends to be at an airmass of 2.6-3.0 depending on time of year.  Observations stop every day at 00:00 UTC (17:00 local time).  Exposures are terminated at an SNR of 500, which on clear days corresponds to an average exposure time of $\sim$180~s (3 minutes).  At an SNR of 500, the analytic RV uncertainty per observation is about 35~\cms.

Solar observations from \expres\ are extracted using largely the same pipeline for nighttime observations \citep{petersburg2020}.  Exposures are calibrated using lines generated from a Menlo Systems laser frequency comb \citep[LFC;][]{wilken2012, molaro2013, probst2014, probst2020, milakovich2020}, which ranges from $\sim490-730$~nm, and a hierarchical, non-parametric wavelength solution \citep{zhao2021}.  RVs are calculated using only the wavelength range for which there is LFC data.  Dark and flat exposures are taken at the beginning of every day.  Exposures are extracted using flat-relative optimal extraction \citep{zechmeister2014}.  ThAr and LFC observations are taken every 30-45 minutes throughout the day.

For each observation, a quality factor is calculated using the standard deviation of the exposure meter counts, $\sigma_\mathrm{EM}$, and the median of the exposure meter counts, $Md_\mathrm{EM}$, as described further in \cite{llama2023}.  The quality factor punishes exposures with large $\sigma_\mathrm{EM}$ values and large $\sigma_\mathrm{EM}/Md_\mathrm{EM}$ ratios.  RVs for each exposure were also regressed against time from solar noon to account for a daily offset thought to be due to the telescope flipping at solar noon (see \cite{llama2023} for more discussion of this issue.).  Other than this regression, no correction is made for the differential atmospheric extinction of the Sun's resolved disk as it sets.

\pagebreak
\subsection{\neid}\label{sec:neid}

The NN-explore Exoplanet Investigations with Doppler spectroscopy (\neid) is a stabilized, fiber-fed, optical to NIR spectrograph with a full wavelength range of $380-1046$~nm spectrograph \citep{schwab2016,NEID_budget}.  It was commissioned at the WIYN 3.5-m telescope
\footnote{The WIYN Observatory is a joint facility of the NSF's National Optical-Infrared Astronomy Research Laboratory, Indiana University, the University of Wisconsin-Madison, Pennsylvania State University, and Purdue University.}
at Kitt Peak National Observatory near Tuscon, Arizona during 2020 and early 2021.  \neid\ has a median resolution of $R\sim120,000$, a total instrumental error budget of 27~\cms\ \citep{NEID_budget}, and an on-sky precision of  41~\cms\ for solar observations \citep{lin2022}. 

The \neid\ Solar Feed is a 75-mm lens with a fiber-feed into \neid\ that has been observing the Sun as a star since December 2020 \citep{lin2022}.  \neid\ starts solar observations every day at 16:31 UTC (9:31 local time) and continues tracking the Sun until 22:30 UTC (15:30 local time) so as not to interfere with the calibration sequence for nighttime observations.  A set of standard calibrations is captured both preceding  and following the solar observations.  Exposures are 55~seconds.  This results in a typical analytic RV uncertainty per observation of $\sim$23~\cms\ across all solar observations (note, the $\sim$16~\cms\ average analytic error cited in Table~\ref{tab:data} is different as that was calculated using just the month of data being analyzed in this work).

Solar observations from \neid\ are extracted using the same pipeline as is used for nighttime observations; this pipeline is  described at \url{https://neid.ipac.caltech.edu/docs/NEID-DRP/}\footnote{All data from the \neid\ solar feed is made available in a timely manner at \url{https://neid.ipac.caltech.edu/search_solar.php}}.  For the RVs used here, the \espresso\ G2 mask is used to calculate RVs, meaning only lines within the wavelength range of the mask ($380-785$~nm) are used to derived the \neid\ DRP RVs.  Exposures are calibrated using a combination of simultaneous observations of a Fabry-Perot etalon, which ranges from $440-930$~nm, and non-simultaneous observations of a Menlo Systems LFC, which ranges from $\sim500-930$~nm. 

The \neid\ calibration protocols were developed for nighttime observations.  In particular, a liquid nitrogen dewar is filled each morning, shortly after starting solar observations.  \neid\ solar observations start at 16:31 UT; the liquid nitrogen refill is triggered at 17 UT.  The dewar is filled in the morning so that any thermal transients, which can cause the instrument to experience large rapid drifts of $\sim$2~\ms, will be settled during nighttime observations.  In contrast, the timing means the instrument drift has not settled during the initial part of the solar observations.  The drift is tracked by the simultaneous calibration and is mostly, but not entirely removed by the DRP.  Many of the solar observations taken at the beginning of a day are therefore affected by nonlinearities in the wavelength calibration that render the resulting velocities less precise.  Since the drift model was developed and tested for nighttime observations, it is likely that a future iteration of the pipeline could improve the wavelength calibration of the daytime observations.  

To determine good quality data (i.e., data for which the majority of the light in the exposure came from the full, integrated disk of the Sun), additional quality cuts were made on the \neid\ data analyzed here.  These quality cuts made use of the time of day of the observation, data from a pyrheliometer, and the exposure meter counts.  Early observations are cut to avoid the nonlinear behavior that follows the dewar filling; late observations are cut to avoid outliers seen on some days likely due to exposures continuing after the solar feed stops tracking and/or is obstructed.  In addition to the exposure meter, \neid\ uses a pyrheliometer to provide an independent measure of the Sun's irradiance.  The standard deviation and mean of both the pyrheliometer measurements and exposure meter counts are used to assure observations were taken during a time of steady and significant solar flux.  The exposure meter flux is then compared to the pyrheliometer flux to ensure they agree, indicating good pointing during the exposure.

\section{Methods}\label{sec:methods}
DRP observations are binned to a common set of timestamps to allow for a direct comparison between the solar data collected by each instrument.  Using the binned RV time series, we can determine the agreement between instruments both within and across a day (\S\ref{sec:direct}).  We additionally measure day-to-day offsets for each instrument (\S\ref{sec:offset}).

For both the binned RVs and the day-to-day offsets, we assume the provided analytic errors give $1\sigma$ Gaussian errors for each data point and use a Monte Carlo approach to determine empirical errors. The given analytic errors are used to generate 1000 random samples of the observed time series.  For each observation, a new random RV is drawn from a normal distribution with a mean of the original RV value and a sigma of the given analytic error for that observation.  Each of the 1000 randomly generated time series is then binned using the same method as described below.  The scatter in the resulting binned point across all simulations is used as the error for the binned point.

The expected amplitude of various sources of error is given in Table~\ref{tab:error}.  For each observation, we have analytic errors from the corresponding instrument's DRP.  The average analytic error as reported by each instrument's DRP is given in Table~\ref{tab:error} under ``Analytic.''  The average error of the binned RVs for each instrument, estimated as described above, is given in Table~\ref{tab:error} under ``Binned.''

\subsection{Binning}\label{sec:binning}

\begin{table*}[tb]
\scriptsize
\caption{Average error from different sources.  All values are given in units of \cms.}
\label{tab:error}
\begin{center}
\begin{tabular}{l c c l c c c c}
\hline
\hline
Instrument & 
Analytic\textsuperscript{(a)} & 
Binned\textsuperscript{(b)} & 
Compared With & 
No. Points\textsuperscript{(c)} & 
Binning RMS\textsuperscript{(d)} & 
Residual RMS\textsuperscript{(e)} & 
Remaining Scatter \\
\hline
\harpsn & 29 & 21 & & & & & \\ 
 &  &  & \harps & 161 & 11 & 36 & 17 \\ 
 &  &  & \expres & 91 & 26 & 47 & 33 \\ 
 &  &  & \neid & 10 & 10 & 20 & - \\ 
\harps & 30 & 9 & & & & & \\ 
 &  &  & \expres & 306 & 25 & 43 & 26 \\ 
 &  &  & \neid & 229 & 14 & 26 & 17 \\ 
\expres & 19 & 12 & & & & & \\ 
 &  &  & \neid & 508 & 26 & 39 & 16 \\ 
\neid & 16 & 6 & & & & & \\ 
\hline
\end{tabular}
\end{center}
{\raggedright (a) average analytic error over the month of shared data; (b) average empirical error of binned RVs; (c) total number of overlapping points; (d) average scatter expected from imperfect binning over p-modes; (e) scatter from the residuals of binned RVs \par}
\end{table*}

The RVs collected by each instrument are interpolated onto shared timestamps to allow for a direct comparison.  The Sun is expected to exhibit p-mode oscillations with a $\nu_\mathrm{max}$ of 5.4 minutes (3.1 mHz), which introduces RV variations on that time scale \citep{kjeldsen2008}.  The different timestamps and observing cadence of the four instruments will therefore be sampling this high frequency signal at different points.  The variation in exposure times used by the different instruments will also average out this signal to different degrees.

We ran tests on simulated p-mode oscillation time series to compare different methods of binning and to estimate the expected error from the binning process.  Oscillations are simulated as a series of damped driven simple harmonic oscillators (SHO).  We include the contributions from 32 SHOs corresponding to $\ell=0,1$ modes.  More SHOs and higher order modes were found not to contribute significantly to the power spectrum.  At each timestamp in a continuous grid with one-second separations, we solve for the position and velocity of each damped simple harmonic oscillator component, apply a stochastic driving force in the form of a randomized velocity perturbation, and sum the velocity contributions from each oscillator (or ``mode'') to yield a simulated asteroseismic RV.

The characteristic frequencies and amplitudes of the SHOs were set to approximate the observed solar p-mode power spectrum.  Specifically, we use a central peak frequency of $\nu_\mathrm{max}$=3100~$\mu$Hz, a peak separation of $\Delta \nu$=1.3$\times10^{-4}$~Hz, and a Gaussian envelope width of 331~$\mu$Hz, in accordance with the values adopted by \citet{chaplin2019}.  The damping timescale was set to two days, a typical value for solar-like stars \citep{chaplin2009}. The amplitude of the stochastic driving force was set to produce a coherent RV amplitude of $A_{\mathrm{max}} = 0.19$ \cms, again in accordance with \citet{chaplin2019}.

These simulated p-modes are then turned into realistic, observed time series that mirror the specifics of the data sets from each instrument.  Observations are derived from the simulated p-modes by first interpolating between the start and end time of an observation onto a finer, 0.1~s grid.  This finely sampled signal is averaged to produce an ``observed'' RV from the simulated p-mode time series.  The time stamp and start/end time of each simulated ``observed'' RV exactly match the properties of the real observations from each instrument.  In this way, the simulated p-modes are converted into observations that match the properties of the real observations being analyzed.

We generated 100 realizations of ``observed'' time series from simulated p-mode oscillations to test a variety of binning methods.  Binning methods varied in time between each shared timestamp (between 5 and 60 minutes) as well as the width of each bin (between 2.5 and 90 minutes) where the bin width was constrained to be greater or equal to the time between each timestamp.  The data in each bin were combined via a weighted average where the weights were determined using either (1) just each observation's given analytic error, or (2) each observation's error as well as the separation in time between the observation and the center of the bin.

Errors for each binned data point were derived using a Monte Carlo (MC) method.  For 1000 simulations, the true DRP observed RVs were perturbed by their given $1\sigma$ analytic errors (the average analytic error is given in Table~\ref{tab:data} and repeated in Table~\ref{tab:error} under the ``Analytic'' column).  These perturbed points were then binned.  The resultant scatter in the binned points across all 1000 simulations was taken to be the MC-derived error for the binned points.  This error is give in Table~\ref{tab:error} under the ``Binned'' column.

\begin{figure}[bhtp]
\centering
\includegraphics[width=0.48\textwidth]{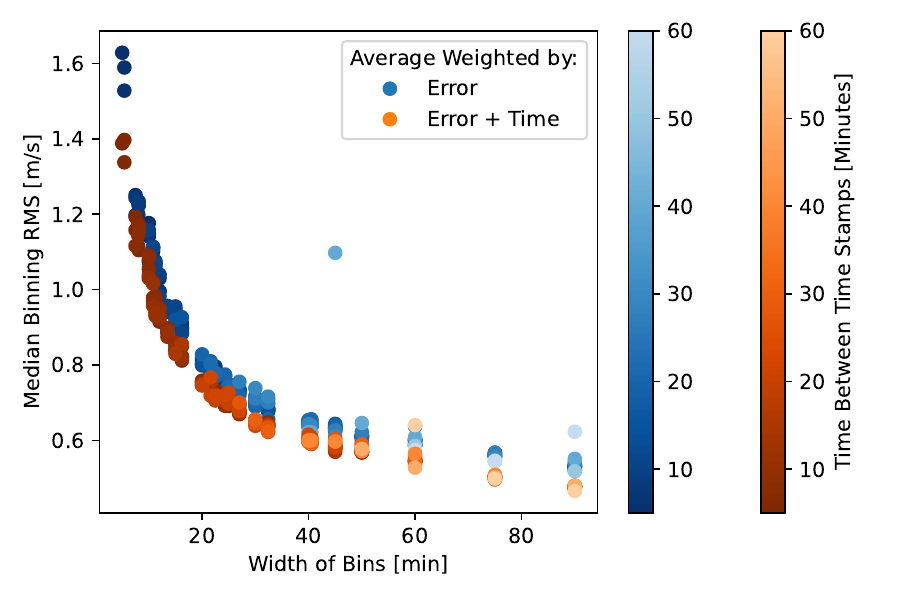}
\caption{Resulting Binning RMS from different binning methods as a function of bin width.  The color of the point indicates whether the RVs in a bin were weighted by analytic error before being averaged (blue) or weighted by error and time from the center of the bin (orange).  The saturation of the color indicates the time between each bin.}
\label{fig:binning}
\end{figure}

Since the point of the binning is to allow for a direct comparison of the solar data gathered by each instrument, the binning methods were evaluated on the scatter returned by the resulting binned RVs when a pair of instruments were differenced.  We take the spread in residuals from these comparisons of simulated observations as the expected error due to the binning process alone; these values are given in Table~\ref{tab:error} under ``Binning RMS.''  The median Binning RMS across the simulated 100 p-mode time series from the different binning methods is shown in Figure~\ref{fig:binning} as a function of bin width and colored by bin spacing and weighting method.

The resultant ``Binning RMS" decreases with the width of each bin following a power law.  This is expected as wider bins allow for more data to be averaged into a shared point.  The resultant RMS values are consistently $\sim$5~\cms\ lower when RVs are weighted by RV error and time of observation (as opposed to just RV error).  The shape of the power law is dominated by the width of the RV bins and is unaffected by the spacing of the time bins.  Fitting the relation to a power law reveals no significant deviations that would suggest a given binning method is doing better than what is expected when more data is added.

\begin{figure}[tp]
\centering
\includegraphics[width=0.48\textwidth]{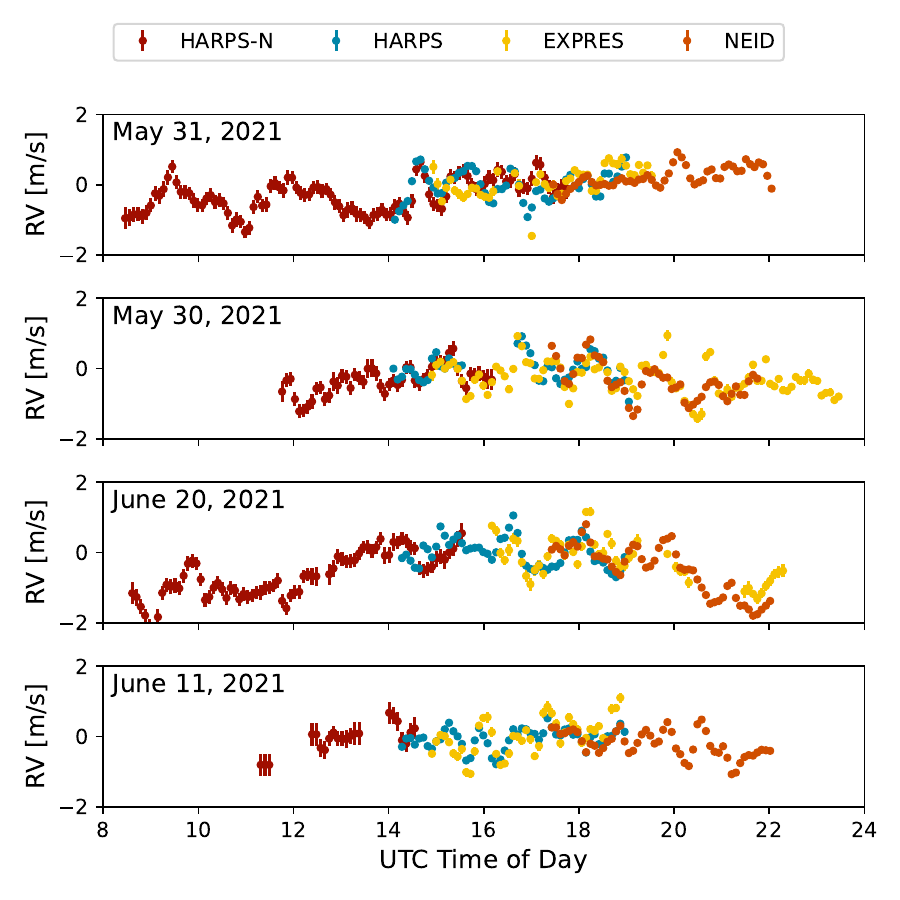}
\caption{Binned RVs as a function of UTC time of day.  We show here the four days for which there was the greatest number of overlapping points between all instruments.  Error bars for each binned RV are derived empirically, as described in Section~\ref{sec:methods}.  A different RV offset is applied for each instrument for each day; the offset is calculated as the median of the binned RVs for that instrument taken on the day in question.  Features that are traced by more than one instrument are most likely solar in nature rather than due to instrument-specific systematics.}
\label{fig:days}
\end{figure}

Because there was no clear evidence of a given binning method returning smaller scatter than expected, we chose a binning method based on what is understood of the Sun and the data used in this analysis. We establish timestamps that are spaced 5.4~minutes apart, the \vmax\ for the Sun.  For each timestamp, all observations within a ($5.4\times3$=)16.2-minute window are combined via a weighted average.  RVs are weighted by both the inverse variance of the observation and a weighting following off linear from the time from the center of the bin.  For example, a point falling exactly at the center of the bin would get a relative weighting of 1 and a point at the edge of the bin (i.e.\ $16.2/2=8.1$ minutes away) would get a relative weighting of zero.  The longer, 16.2-minute window ensures that there are enough observations in a bin to produce a reasonable average, even for observations with the longest exposure times (5+ minutes).  Some observations will be included in multiple bins, resulting in a smooth binned behavior as expected.  Binned RVs are shown in Figure~\ref{fig:days} for the days on which there is the greatest amount of overlap between the four instruments.

\begin{figure*}[bt]
\includegraphics[width=\textwidth]{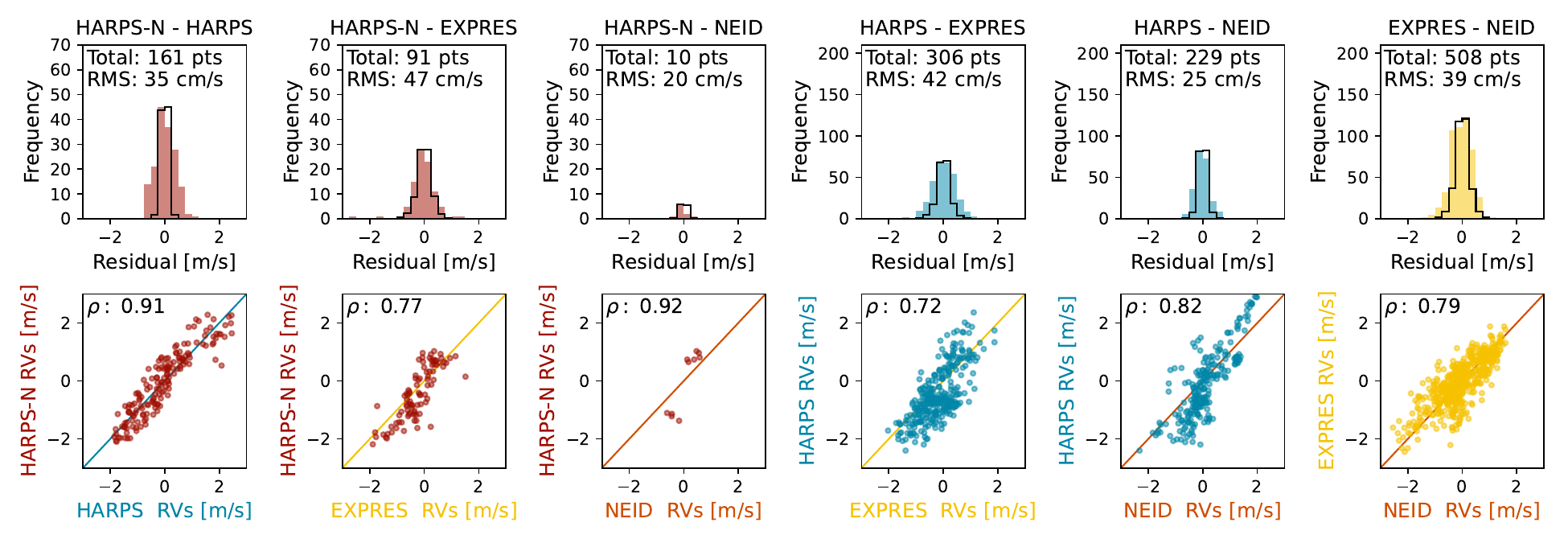}
\caption{\textbf{Top}: Histograms of the residuals between one instrument's binned RVs against another instrument's binned RVs.  The title of each subplot details the two instruments being compared (where the binned RVs of the second instrument listed is subtracted from the binned RVs of the first instrument).  The total number of overlapping binned RVs across the entire month of shared data is given at the top of each subplot followed by the RMS of the residuals.  \textbf{Bottom}: Comparison of one instrument's binned RVs (y-axis) against the binned RVs of the other three instruments (x-axis).  A point is shown for each binned timestamp at which both instruments have a binned RV.  The corresponding Pearson correlation coefficient, $\rho$, for each pairwise comparison is given in the top-left corner.  Values of $\rho$ closer to 1 indicate more significantly positively correlated data.}
\label{fig:residual_corr}
\end{figure*}

\subsection{Direct Comparison of Binned RVs}\label{sec:direct}
Binning observations from each instrument to shared timestamps allows us to directly compare binned RVs between instruments.  Before comparison, an RV offset, calculated as the median of all binned RVs, is removed from each time series.  Figure~\ref{fig:residual_corr} shows the spread in the residuals when the binned RVs from different pairs of instruments are differenced.  Each of the six columns corresponds to a different pair of instruments, as given at the top of each column, where the total number of overlapping points between the two instruments is written at the top of the plot.

This spread in residuals, as written on the top of each subplot, is also given in the table of errors (Table~\ref{tab:error}) under ``Residual RMS''.  The distribution of residuals from the simulated p-mode oscillation is overplotted in black.  In other words, the expected spread from just the error introduced from binning is shown by the black outline histograms.

The bottom row of the figure shows the binned RVs from each instrument plotted directly against the binned RVs from each of the other instruments.  The Pearson correlation coefficient, $\rho$, between the two binned RV time series is given in the top-left of each subplot.

The last column of Table~\ref{tab:error} gives the magnitude of the remaining unexplained scatter.  As described, we have quantified an error due to (1) photon noise, as given by the analytic error for each data point, and (2) the binning process, as determined using simulated p-mode oscillation time series.  Assuming these two noise sources are independent, Gaussian, and make for a complete accounting of the noise present in the comparison, then we would expect the residual RMS to be the analytic error and the binning RMS added in quadrature.  In actuality, we find excess noise in the residual RMS, which is given in the ``Remaining Scatter'' column of Table~\ref{tab:error} (i.e., this is the amount of error that still has to be added in quadrature, in addition to the analytic and binning error, to obtain the residual RMS).  For the comparison of \harpsn\ with \neid, no remaining scatter value is given because there are too few data points (only 11).

\begin{figure}[thp]
\centering
\includegraphics[width=0.5\textwidth]{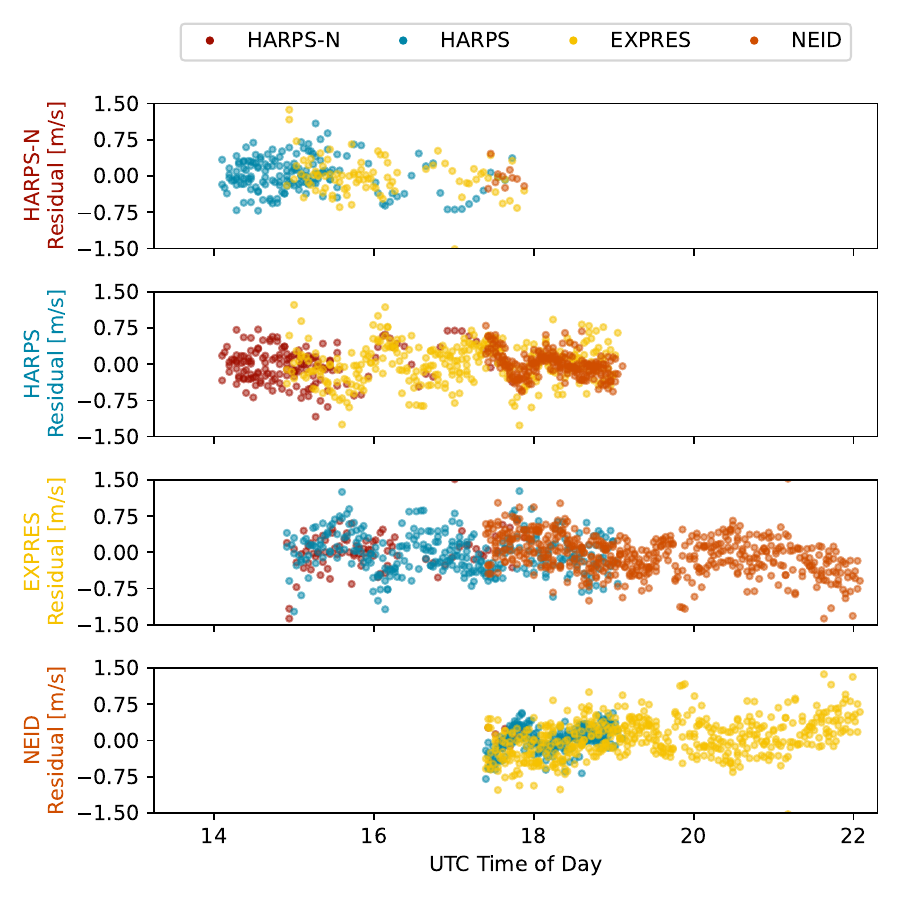}
\caption{Residuals between binned RVs across all 29 days phase-folded to UTC time of day.  Each subplot shows the residuals between binned RVs of the instrument specified in the y-axis label and the other three instruments.  A point is shown for each binned timestamp at which both instruments have a binned RV.  Of particular interest are features traced by the residuals from multiple instruments, which lends confidence to the feature and helps diagnose its origin.}
\label{fig:residual_tod}
\end{figure}

Figure~\ref{fig:residual_tod} shows pairwise residuals for each instrument phase-folded to UTC time of day.  For each of the four instruments, the residuals with respect to all three of the other instruments are shown.  Phase-folding the residuals to time of day is useful to emphasize non-astrophysical effects.  We would not expect the Sun to exhibit any coherent variation with respect to time of day on Earth.  Therefore, any coherent structures seen as a function of time of day is likely due to Earth-bound observing conditions or daily data processing effects.

Features seen in the residuals with respect to more than one instrument are of particular interest.  Having the same coherent variation appear in more than one comparison provides independent confirmation of the variation.  Additionally, with pairwise comparisons, having the same features appear in multiple comparisons is needed to pinpoint the origin of the feature.  As an example, if a comparison between instruments A and B reveal a coherent variation, it is unclear if it is instrument A, instrument B, or a combination of the two causing that coherent variation in the absence of ground truth.  However, if a comparison between instruments A and C exhibits the same coherent variation, then we are able to conclude that instrument A is the origin of the variation.

\subsection{Day-to-Day Offsets}\label{sec:offset}

We measure day-to-day offsets (i.e., a single RV offset for each day of solar data) for each instrument relative to the other instruments.  We know there exists measured RV variations due to changes on the solar surface.  Ideally, it would be possible to construct a model of these variations.  The RVs from each instrument for each day could then be compared to this model, giving an offset for each day.  With the current data, it was not possible to get a good fit for such a model.  Because there is a minimum eight-hour gap in solar data every day (between when the Sun set for the west-most instrument and when the Sun rose again for the east-most instrument), there is little constraint on the model from one day to the next.


Rather than prescribe a potentially flawed model of the expected RV variations, we compare each instrument to the RVs measured by the other instruments.  Because all instruments are observing the same solar surface variations, it is reasonable to use the stellar signals measured by another instrument in place of a model of the expected RV variations.  This, of course, ignores the inherent error of each observation from photon noise, instrument systematics, and varied timestamps/exposure times.

\begin{figure}[tp]
\centering
\includegraphics[width=.48\textwidth]{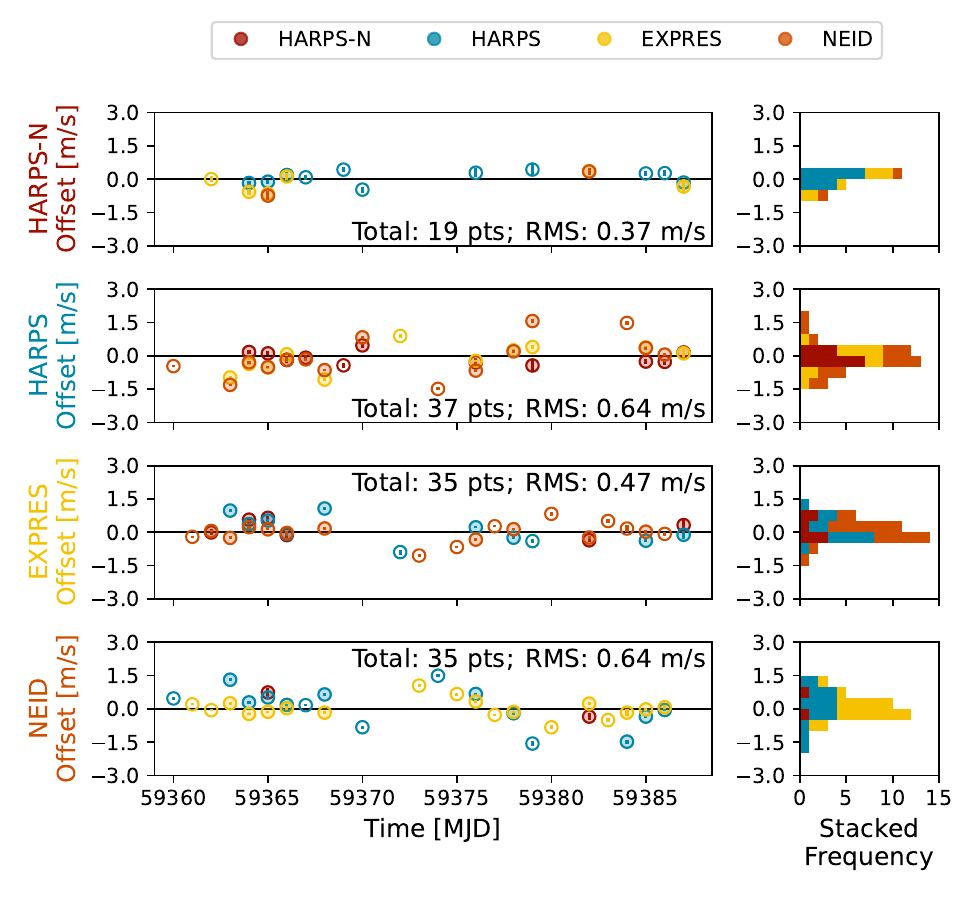}
\caption{\textbf{Left}: Day-to-day offsets relative to other instruments.  Each subplot shows the day-to-day offsets for the instrument specified in the y-axis label as calculated relative to the other three instruments.  A point is shown for each day in which both instruments have at least one overlapping binned RV.  Days for which there exist offsets measured against two or more instruments are shown as filled-in circular points; days for which there only exists one comparison are shown as outlines.  Note: the expected error due to binning for each point is shown as an errorbar, but all errorbars lay interior to each point suggesting that binning error is not the dominant source of uncertainty.  The total number of days for which an offset was calculated across all three comparisons as well as the RMS of all offsets is written on the right of each subplot.  \textbf{Right}: Stacked, horizontal histograms of the relative day-to-day offsets calculated for each instrument.  The resulting distribution of comparisons against each of the three instruments is stacked on top of each other.  In other words, the height of a multicolored bar is the frequency of that bin's values across the comparison with multiple instruments.}
\label{fig:daily_off}
\end{figure}

To measure the offset between two instruments over a day, we find the difference between the median of the binned RVs, where only the binned RVs where the two instruments overlap are used to calculate this median.  The resultant day-to-day offsets are shown for each instrument in Figure~\ref{fig:daily_off}.  Points are only shown for days where the two instruments had overlapping binned RVs.  Note, however, that the number of overlapping points differs between instruments and between days, meaning the offsets are not directly comparable.  Offsets on days for which there exists offsets calculated relative to two or more instruments are shown as filled-in points.  Of particular interest are days on which multiple offset calculations are consistent with one another.  Measurements from different comparisons that agree lends confidence to the measured offset for the instrument in common between the comparisons.

The error bars in Figure~\ref{fig:daily_off} are derived using the simulated p-mode time series described above in Section~\ref{sec:binning}.  The resultant observed time series derived from the simulated p-modes are used to calculate day-to-day offsets for each instrument as described.  The spread in the returned offset across all 1000 simulations is shown as the error in Figure~\ref{fig:daily_off}.  Note, this error only reflects the expected error from binning p-modes into discrete observations and then to shared time stamps.  It does not capture errors that would, for example, arise from instrument systematics, varied levels of differential extinction, or any other, non-binning related source of uncertainty.

\section{Results}\label{sec:results}
We are able to probe instrument-to-instrument precision within a day, across a day, and across the 29 days of shared data.  Precision within a day measures the degree to which each instrument returns the same RV variation in response to solar surface changes.  Should the agreement between instruments vary across a day, this may imply day-long instrument drifts or imperfect correction for daily effects, such as atmospheric extinction or instrument drift.  Instrument precision and accuracy accuracy from day-to-day hints at the long-term stability of each instrument and, consequently, an instrument's ability to recover long-period signals.

\subsection{Precision Within a Day}

RVs from multiple instruments often trace the same features.  It is easy to visually see this in Figure~\ref{fig:days} where the binned RVs from each instrument are shown for the four days on which there is the greatest amount of overlap.  Though the binning cannot completely mitigate the differences in RV that arise simply from different timestamps, we see here that the binned RVs from each instrument do seem to largely agree on the resultant measured RV variation from solar surface changes.

While the bin widths average over p-mode oscillations \citep{chaplin2019}, there remains short-timescale oscillations in the binned RVs likely due to granulation.  Given that the same oscillatory features are traced by multiple instruments, they are likely astrophysical in nature.  Similar variations have been found in \harps\ astroseismology measurements and in an analysis of three years of \harpsn\ solar data \citep{dumusque2011-01, collier2019}.  These analyses concluded that such short-timescale variations are likely due to granulation phenomena.

Histograms of the residuals between the binned RVs of pairs of instruments (top row of Figure~\ref{fig:residual_corr}) show the agreement between all overlapping binned RVs.  The expected error from comparing binned RVs between instruments is given in Table~\ref{tab:error} in the ``Binning RMS'' column.  Comparisons to the \expres\ binned RVs give rise to the largest expected errors.  Of the four instruments, \expres\ is the only instrument that makes use of adaptive exposure times, exposing to a set SNR rather than a set time.  Since we expect the error from binning to scale directly with exposure time, having changing exposure times may change the expected binning error throughout the \expres\ time series, leading to these larger ``Binning RMS'' values.

After taking into account the expected error from generating binned RVs to the shared timesteps, we find that the pairs of instruments agree with each other to within 15-30~\cms\ (see the ``Remaining Scatter'' column of Table~\ref{tab:error}).  This shows that even using the default DRP derived RVs, the instruments are in very good agreement with one another.  The bottom row of Figure~\ref{fig:residual_corr} plots the binned RVs from the two instruments against each other.  The RVs are linearly correlated and have significant correlation coefficients in all cases as expected.

\subsection{Precision Across a Day}

To investigate variations in instrument precision across a day, we show the residuals for each instrument phase-folded to UTC time of day in Figure~\ref{fig:residual_tod}.  Any coherent structure as a function of time of day is most likely a cause of necessary daily corrections---e.g., to account for things such as instrument drift, or Sun-specific effects, such as the changing atmospheric extinction as the Sun rises and sets, etc.---rather than true variations on the solar surface.

There appear to be oscillations in the \harps\ residuals (see the second sub-plot from the top of Figure~\ref{fig:residual_tod}.  These oscillations have a semi-major amplitude of ~85 \cms\ and a period of 30-50 minutes.  The oscillations show most clearly in the residuals as compared to \expres\ data (\exprescolor\ points), for which there is the greatest amount of overlap with the \harps\ data, but the same oscillations also appear in the residuals to the \neid\ RVs (\neidcolor) and possibly the \harpsn\ residuals (\harpsncolor) as well.  We can therefore attribute this behavior to the \harps\ measurements.  This feature is only seen during a short period of the \harps\ solar observations, including the period analyzed in this paper.  It is likely due to an unusual amount of dust on the telescope transparent dome that induced differential extinction.  This effect will be investigate further in a forthcoming paper.

There appears to be a slope in the residuals between the \expres\ and \neid\ RVs when the Sun would be setting for these two instruments.  After UTC time of day $\sim$21, \neid\ RVs are increasingly higher, or more red-shifted, than \expres\ RVs as the Sun sets.  The slope of this trend is approximately 73~\cms\ per hour as fit using data taken after a UTC time of 21, where the slope is the steepest.  This may indicate a need to account for high airmass effects or differential extinction.  Though neither instrument explicitly accounts for differential extinction, the solar noon correction applied to the \expres\ observations (as described in Section~\ref{sec:expres}) likely mitigate differential extinction to some degree.  \harps\ and \harpsn\ do explicitly account for differential extinction (as described in Sections~\ref{sec:harps} and \ref{sec:harpsn}) and do not show a similar linear trend.  This trend can therefore in principle be attributed to whether or not the effects of high airmass and differential extinction are corrected for.

\subsection{Stability Across Days}

RVs from multiple instruments often trace the same long-term features as well as daily ones.  Figure~\ref{fig:rvs} plots the binned RVs from all instruments in the bottom-most subplot.  The RVs from each instrument generally follow the same trend throughout the month.

Figure~\ref{fig:daily_off} shows the derived day-to-day offsets for each instrument relative to the other instruments.  We chose to measure offsets relative to the other instruments to avoid introducing any potential errors from assigning a flawed model of the expected measured RV variations.  This pairwise comparison, however, comes with its own caveats.

The measured daily offsets are derived using variable amounts of data.  An offset is shown for any day on which there exists overlapping RVs for two instruments.  The number of binned RVs used to calculate the offset therefore varies from a single pair of overlapping binned RVs to nearly 60 overlapping points in some cases.  The errorbar shown for each point, which are on the order of 5-25 \cms, captures only the expected error from binning and so does not reflect the different amounts of information going into each of the points.  Because of the varied amounts of overlapping data from day to day, it is difficult to directly compare the measured offsets with one another as they differ widely in information content.

There do exist some days for which the day-to-day offset for an instrument is reflected in the comparison against more than one instrument.  Because all offsets are relative and pairwise, it is hard to tell from a single comparison what the source of the offset is.  If comparisons against two different instruments return the same offset, however, that lends credibility to the returned offset.  We therefore emphasize days for which an instrument has a calculated offset relative to two or more instruments by showing these days with filled-in circular markers in Figure~\ref{fig:daily_off}.

Figure~\ref{fig:daily_off} shows stacked histograms to visualize the distribution of the day-to-day offsets calculated for each instrument.  The histograms are stacked to show the overall distribution of all calculated offsets.  The color of the bars correspond to the instrument used to calculate each day-to-day offset.  This makes it easier to determine if any skewness in the distribution is due to one instrument in particular.  For example, the \neidcolor-colored bars in the third plot represent day-to-day offsets for \expres\ as compared to \neid\ data.  The \neidcolor\ bars are primarily responsible for the stacked histogram appearing to be skewed towards positive offsets.  This matches our finding from Figure~\ref{fig:residual_tod} that residuals between \expres\ and \neid\ binned RVs are skewed positive as the Sun sets.

We note that that this way of calculating the day-to-day offset is particularly sensitive to differential extinction when comparing instruments in different time zones.  The effects of differential extinction are not only greatest when the Sun is rising or setting, the effect is also opposite between the Sun rising and the Sun setting.  For instruments with a large separation in time zones, many of the overlapping data points will be when the Sun is setting in one location and rising in the other.  The spread in the daily offsets between \harpsn\ as compared to \expres\ or \neid\ (shown by the \exprescolor\ and \neidcolor\ bars in the second row of the right-column of Figure~\ref{fig:daily_off}) is larger than the spread in the daily offsets between \expres\ and \neid.  This could be because \expres\ and \neid\ are geographically close, and therefore view the Sun concurrently at a similar airmass, while \harpsn\ is further to the east and south.

We find that the scatter in day-to-day offset for each instrument is on the order of 50-60~\cms.  This is likely artificially inflated given the shortcomings of using relative offsets as described above.  More accurate estimates can be achieved with data that more completely spans the 24 hours in a day and more days than the 29 days used in this analysis.

\section{Discussion}\label{sec:discussion}
Combining solar data from different instruments makes for a powerful test bed on which to understand methods for mitigating stellar signals.  With solar data, it is possible to confidently remove measured velocity shifts due to Solar System bodies.  Because these solar feeds collect sunlight in a way that closely emulates light collection of other stars, the resultant data set captures real stellar surface variations in the same manner as would  be observed on other stars.  High SNR and high cadence observations are uniquely easy to obtain, and it is simple to degrade these observations to approximate the lower-SNR and sparser observations of nighttime targets.  By combining observations from multiple instruments, we are also able to probe systematics that are unique to individual instruments.

There are also limitations of observing the Sun.  As a resolved source, differential extinction as the Sun rises and sets causes spurious RV shifts at high airmass, something that does not happen to other stars, which appear as point sources \citep{davies2014, collier2019}.  Because the Earth orbits the Sun, the barycentric corrections for solar data is of a much smaller magnitude than when observing other stars.  Telluric lines therefore shift with respect to stellar lines to a lesser degree, making it far more difficult for methods that depend on a large difference in stellar and barycentric shift to be tested on solar data \citep[e.g.][]{bedell2019,cretignier2021}.  Solar light is often injected into the fiber leading to the spectrograph via a different optical path than nighttime observations, which are taken with the main telescope.  Analyses using solar data are therefore insensitive to possible instrument systematics unique to nighttime observing, such as the guiding and focusing of the main telescope.  Lastly, the Sun is just one example of a star, and how its surface variations map to the surfaces of other stars with different spectral types and stellar properties has not been well characterized.

\subsection{Standardizing RV Derivation}

Here we use the default pipeline RVs from each instrument and find that the different instruments agree very well with each other.  The binned RVs from each instrument are strongly correlated with other instruments and can be visually seen to trace similar patterns.  When taking into account the given analytic error and the error from binning, we find that a direct comparison between RVs from different instruments show that they agree to within 15-30~\cms.

It might be possible to get even better agreement between instruments by standardizing how the RVs are derived.  \harps, \harpsn, and \neid\ all produce RVs using the cross-correlation function (CCF), where a mask is shifted to find the best match with the cores of a given list of spectral lines.  The \expres\ RVs are derived using a forward-modeling framework that uses all spectral lines in the range where there is LFC light ($\sim490-730$~nm).  The contribution of each line to the final RV is based on an empirical measure of the RV content of the line.

We expect different lines to respond differently to stellar signals.  Therefore, we would expect RVs to agree the best when the same stellar lines are used to derive them.  All CCF RVs used in this analysis were derived using the \espresso\ G2 line list, and so should have incorporated many of the same lines across a similar wavelength range.  We also find that the blaze function of each instrument peaks at similar locations for all instruments (see Figure~\ref{fig:throughput}), meaning the relative weighting of a spectral line across an order will be similar for all instruments.

Despite similar line lists and throughput across an \'echelle order, there still exists many ways in which the pipeline RVs differ.  \expres\ RVs are derived using a forward-modeling method that only incorporates wavelengths for which there is LFC coverage ($\sim$490-730~nm).  Even for \harps, \harpsn, and \neid, which all use the CCF method and the \espresso\ G2 mask (with lines between $\sim$380-785~nm) for deriving RVs, the CCF is calculated by different pipelines.  

Different pipelines also differ in the amount of information per order and the way the RV information content from each order is combined.  The size of the detector differs between instruments, changing the wavelength range of each individual \'echelle order and, consequently, the number of spectral lines that appear in more than one order.  The overall throughput as a function of wavelength also varies by instrument.  The relative signal of a line in the blue as compared to a redder line therefore changes with instrument.  The same CCF mask is used by \harps, \harpsn, and \neid, to derive a CCF for each order, but each order's CCF/RV is weighted and combined into a global RV for an observation in a different way by the different pipelines.

The different pipelines avoid telluric contamination in different ways.  \expres\ uses a telluric model derived by \selenite\ \citep[a SELf-calibrating, Empirical, Light-Weight liNear regressIon TElluric model;][]{leet2019} to identify and mask out telluric lines.  Though \harps, \harpsn, and \neid\ all use the \espresso\ G2 CCF mask, the different pipelines have different cutoffs for avoiding shifting line masks over telluric lines.

It may therefore be instructive to run the same exact pipeline on solar data collected by different instruments in order to eliminate pipeline differences and isolate stellar and instrumental effects.  Because we find that the pipeline RVs are themselves in fairly good agreement, we leave investigating potential improvements by standardizing RV derivation to future work.

\hfill
\subsection{Benchmarking Instruments Against Instruments}

Concurrent observations of the same source from four different instruments allowed us to benchmark instruments against each other and gain a deeper understanding of the observations.  Concurrent observations removed the need for constructing potentially flawed models to compare against.  However, the different timestamp and exposure times of observations introduced errors in how high-frequency signals were sampled within each time series.  This would not be an issue for longer-period signals (e.g., on the time scale of planet periods or stellar rotation rates).  Instrument-to-instrument comparisons were also only possible when concurrent observations fell very close to each other; data with observational gaps and/or telescopes geographically far apart reduce the amount of overlapping data that can be used to inform a rigorous comparison.

The differential extinction of integrated sunlight as the Sun rises and sets further complicates comparisons between instruments.  For geographically distant instruments, much of the overlapping data will be taken as the Sun is setting in one location and rising in another.  Comparisons between instruments will then mostly use observations taken at high airmass and subject to the effects of differential extinction in opposite ways (i.e., the difference between the effect as the Sun rises vs.\ sets).  This is likely to artificially inflate the scatter in the residuals and calculated day-to-day offsets.

We find the spread in the day-to-day offset for each instrument is on the order of 50-60~\cms, which is much larger than the intra-day residual scatter of 15-30~\cms.  It is possible that instrument systematics or data reduction artifacts truly cause the RVs from each instrument to change by up to 60~\cms\ from day to day.  However, it is more likely that the calculated day-to-day offsets are artificially inflated due to differential extinction, which can cause an RV difference of about 50~\cms\ at an airmass of 2.2 \citep{collier2019}, an effect that will compound if observations taken when the Sun is setting are compared to observations taken by a different telescope when the Sun is rising.  Differences may also be mitigated through more standardized pipelines or observing strategies.  It will be difficult to reach a significant conclusion about the day-to-day offsets for different instruments without more overlapping data that is taken with comparable time sampling.

Comparing solar telescopes in the same time zone allows for more confident comparisons.  In this analysis, \expres\ and \neid\ are in the same time zone and share by far the greatest number of overlapping data points.  It will soon be possible to perform a similar analysis with \harps\ and \harpsn\ data as solar telescopes for \espresso\ \citep[\poet;][]{leite2022}\footnote{See \url{http://poet.iastro.pt} for more information} and \harps-3 \citep[\aboras;][]{aboras} come online, covering the same time range as \harps\ and \harpsn\ respectively.

Wider time coverage of solar data will also allow for a better estimate of day-to-day offsets.  There exists global, ground-based networks of solar observatories, such as the Global Oscillation Network Group \citep[\gong;][]{harvey1996} and the Birmingham Solar Oscillations Network \citep[\bison;][]{hale2016,davies2014}.  It is worth exploring if the single-line spectra produced by these networks can provide insight into the expected RV variations as measured by the solar telescopes investigated here.  The space-based Solar Dynamics Observatory \citep[SDO;][]{pesnell2012} also provides near continuous images of the Sun at a variety of wavelengths as well as a measure of the Sun's magnetic activity, which could be used to inform the expected RV response from solar surface variation \citep[e.g.][]{ervin2022}.  It is also possible, in principle, to collect solar data at night via reflected light from the Moon or asteroids to increase overlap between instruments.

This analysis is imminently repeatable and is guaranteed to only get more interesting.  Over the next few years, several new solar feeds for precision spectrographs are expected to come online.  \pepsi\ \citep{strassmeier2018} and \gianob\ \citep{claudi2018} also have solar feeds.  Solar feeds for KPF \citep{rubenzahl2023}, \espresso\ \citep{leite2022}, \harps-3 \citep{aboras}, \maroonx\ \citep{seifahrt2018}, and \nirps\ \citep{bouchy2017} are in the works.  We note that shorter exposure times give rise to less error when binned to share time stamps, meaning shorter solar exposures will be easier to compare with observations from other instruments.

The observations themselves will also become more interesting as the Sun moves out of an activity minimum.  The data used in this analysis were taken only 14 months after the latest cycle minimum, which occurred in December 2019.  Activity on the Sun is expected to peak late in the Fall of 2024 with a maximum sunspot number of 134$\pm$8 \citep{upton2023}.

It is also, of course, possible to perform a similar analysis with observations of stars other than the Sun.  Observations of other stars are less susceptible to differences in airmass and do not experience differential extinction as solar observations do.  This would remove an error source in bench marking instrument variations that was encountered in this analysis of solar data.  Repeating such a comparison with more stars will also inform instrumental and stellar variation across different stellar properties.

From this analysis, it is clear that coordinating observation times will have a huge effect on the ability of different data sets to benchmark one another.  Shorter observations are more easily binned to shared timestamps with less inherent error from binning over high-frequency variations.  Given the presence of short-timescale variations, observations from different instruments should be taken as close in time to each other as possible.  Therefore, targets that are bright and high in the sky will be easiest to observe in a way that is favorable for cross-instrument comparisons.  As such coordination is sure to be logistically difficult, the consistency and high SNR of solar observations will continue to remain valuable even as cross-instrument analyses are pursued with other stars.

By combining solar data sets from different instruments, we were able to learn more about individual instrument systematics, confirm observations of solar signals, and produce a powerful data set for gauging the performance of different methods for mitigating stellar signals.  We find that binned RVs from each instrument all agree to within 15-30~\cms\ when analytic and binning errors are taken into account, even though the four instruments started operations almost 20 years apart.  This demonstrates the precision that the four instruments analyzed here are capable of, showing that small-amplitude signals can be confidently detected with existing instrumentation if stellar signals are successfully disentangled to the needed levels.  This analysis showcases the potential of such a comparison, which is repeatable for a larger number of instruments, a longer time baseline of data, and/or even for stars other than the Sun.

\acknowledgements

We are very grateful to the work of Thibault Pirson, that helped in designing and integrating HELIOS as part of a master project at the University of Geneva.

The HARPS-N project was funded by the Prodex Program of the Swiss Space Office (SSO), the Harvard University Origin of Life Initiative (HUOLI), the Scottish Universities Physics Alliance (SUPA), the University of Geneva, the Smithsonian Astrophysical Observatory (SAO), the Italian National Astrophysical Institute (INAF), University of St. Andrews, Queen’s University Belfast, and University of Edinburgh.

These results made use of the Lowell Discovery Telescope at Lowell Observatory. Lowell is a private, non-profit institution dedicated to astrophysical research and public appreciation of astronomy and operates the \ldt\ in partnership with Boston University, the University of Maryland, the University of Toledo, Northern Arizona University and Yale University.

The \expres\ team acknowledges support for the design and construction of \expres\ from NSF MRI-1429365, NSF ATI-1509436 and Yale University. DAF gratefully acknowledges support to carry out this research from NSF 2009528, NSF 1616086, NSF AST-2009528, the Heising-Simons Foundation, and an anonymous donor in the Yale alumni community. 

S.M.\ is the \neid\ Principal Investigator.
J.T.W.\ and P.R.\ serve as \neid\ Instrument Team Project Scientists.  
F.H.\ is the \neid\ Project Manager.
This paper contains data taken with the \neid\ instrument, which was funded by the NASA-NSF Exoplanet Observational Research (NN-EXPLORE) partnership and built by Pennsylvania State University. \neid\ is installed on the WIYN telescope, which is operated by the NSF's National Optical-Infrared Astronomy Research Laboratory (NOIRLab), and the
\neid\ archive is operated by the NASA Exoplanet Science Institute at the California Institute of Technology. Part of this  work was performed for the Jet Propulsion Laboratory, California Institute of Technology, sponsored by the United States Government under the Prime Contract 80NM0018D0004 between Caltech and NASA.
This paper is based in part on observations at Kitt Peak National Observatory, NSF’s NOIRLab, managed by the Association of Universities for Research in Astronomy (AURA) under a cooperative agreement with the National Science Foundation. The authors are honored to be permitted to conduct astronomical research on Iolkam Du’ag (Kitt Peak), a mountain with particular significance to the Tohono O’odham.
We thank the \neid\ Queue Observers and WIYN Observing Associates for their skillful execution of \neid's nighttime observing programs and careful monitoring of \neid's calibration exposures.
Deepest gratitude to Zade Arnold, Joe Davis, Michelle Edwards, John Ehret, Tina Juan, Brian Pisarek, Aaron Rowe, Fred Wortman, the Eastern Area Incident Management Team, and all of the firefighters and air support crew who fought the recent Contreras fire. Against great odds, you saved Kitt Peak National Observatory.

The Center for Exoplanets and Habitable Worlds and the Penn State Extraterrestrial Intelligence Center are supported by Penn State and the Eberly College of Science.

Computations for this research were performed on the Penn State’s Institute for Computational and Data Sciences' Advanced CyberInfrastructure (ICDS-ACI).  This content is solely the responsibility of the authors and does not necessarily represent the views of the Institute for Computational and Data Sciences.

RMR acknowledges support from the Heising-Simons Foundation.

NCS acknowledges Fundes by the European Union (ERC, FIERCE, 101052347). Views and opinions expressed are however those of the author(s) only and do not necessarily reflect those of the European Union or the European Research Council. Neither the European Union nor the granting authority can be held responsible for them. This work was supported by FCT - Fundação para a Ciência e a Tecnologia through national funds and by FEDER through COMPETE2020 - Programa Operacional Competitividade e Internacionalização by these grants: UIDB/04434/2020; UIDP/04434/2020.

GS acknowledges support provided by NASA through the NASA Hubble Fellowship grant HST-HF2-51519.001-A awarded by the Space Telescope Science Institute, which is operated by the Association of Universities for Research in Astronomy, Inc., for NASA, under contract NAS5-26555.

\bibliography{main}

\end{document}

%% file: authors.tex
\correspondingauthor{Lily L.\ Zhao}
\email{lzhao@flatironinstitute.org}


\author[0000-0002-3852-3590]{Lily L.\ Zhao}
\affiliation{Flatiron Institute, Simons Foundation, 162 Fifth Avenue, New York, NY 10010, USA}

\author[0000-0002-9332-2011]{Xavier Dumusque} 
\affiliation{Astronomy Department of the University of Geneva, 51 ch.\ des Maillettes, 1290 Versoix, Switzerland}

\author[0000-0001-6545-639X]{Eric B.\ Ford} 
\affiliation{Department of Astronomy \& Astrophysics, 525 Davey Laboratory, The Pennsylvania State University, University Park, PA, 16802, USA} 
\affiliation{Center for Exoplanets and Habitable Worlds, 525 Davey Laboratory, The Pennsylvania State University, University Park, PA, 16802, USA} 
\affiliation{Institute for Computational \& Data Sciences, The Pennsylvania State University, University Park, PA, 16802, USA}

\author[0000-0003-4450-0368]{Joe Llama} 
\affiliation{Lowell Observatory, 1400 W. Mars Hill Rd., Flagstaff, AZ 86001, USA)}
 
\author[0000-0001-7254-4363]{Annelies Mortier} 
\affiliation{School of Physics \& Astronomy, University of Birmingham, Edgbaston, Birmingham, B15 2TT, UK}
 
\author[0000-0001-9907-7742]{Megan Bedell} 
\affiliation{Center for Computational Astrophysics, Flatiron Institute, Simons Foundation, 162 Fifth Avenue, New York, NY 10010, USA} 


\author[0000-0002-3212-5778]{Khaled Al Moulla} 
\affiliation{Astronomy Department of the University of Geneva, 51 ch.\ des Maillettes, 1290 Versoix, Switzerland}

\author[0000-0003-4384-7220]{Chad F.\ Bender}
\affiliation{Steward Observatory, University of Arizona, 933 N.\ Cherry Ave, Tucson, AZ 85721, USA}

\author[0000-0002-6096-1749]{Cullen H.\ Blake}
\affiliation{Department of Physics and Astronomy, University of Pennsylvania, 209 S 33rd St, Philadelphia, PA 19104, USA}

\author[0000-0002-9873-1471]{John M.\ Brewer}
\affiliation{Dept. of Physics \& Astronomy, San Francisco State University, 1600 Holloway Ave., San Francisco, CA 94132, USA}
 
\author[0000-0002-8863-7828]{Andrew Collier Cameron} 
\affiliation{SUPA, School of Physics \& Astronomy, University of St Andrews, North Haugh, St Andrews, KY169SS, UK}
\affiliation{Centre for Exoplanet Science, University of St Andrews, North Haugh, St Andrews, KY169SS, UK}

\author[0000-0003-1784-1431]{Rosario Cosentino} 
\affiliation{Fundaci\'on Galileo Galilei-INAF, Rambla Jos\'e Ana Fernandez P\'erez 7, 38712 Bre\~na Baja, Tenerife, Spain}

\author[0000-0001-8504-283X]{Pedro Figueira} 
\affiliation{Astronomy Department of the University of Geneva, 51 ch.\ des Maillettes, 1290 Versoix, Switzerland}
\affiliation{Instituto de Astrof\'isica e Ci\^encias do Espa\c{c}o, Universidade do Porto, CAUP, Rua das Estrelas, 4150-762 Porto, Portugal}

\author[0000-0003-2221-0861]{Debra A.\ Fischer}
\affiliation{Department of Astronomy, Yale University, 52 Hillhouse Ave., New Haven, CT 06511, USA}

\author[0000-0003-4702-5152]{Adriano Ghedina} 
\affiliation{Fundaci\'on Galileo Galilei-INAF, Rambla Jos\'e Ana Fernandez P\'erez 7, 38712 Bre\~na Baja, Tenerife, Spain}

\author{Manuel Gonzalez} 
\affiliation{Fundaci\'on Galileo Galilei-INAF, Rambla Jos\'e Ana Fernandez P\'erez 7, 38712 Bre\~na Baja, Tenerife, Spain}

\author[0000-0003-1312-9391]{Samuel Halverson}
\affiliation{Jet Propulsion Laboratory, California Institute of Technology, 4800 Oak Grove Drive, Pasadena, California 91109}

\author[0000-0001-8401-4300]{Shubham Kanodia}
\affiliation{Earth and Planets Laboratory, Carnegie Institution for Science, 5241 Broad Branch Road, NW, Washington, DC 20015, USA}
\affiliation{Department of Astronomy \& Astrophysics, 525 Davey Laboratory, The Pennsylvania State University, University Park, PA, 16802, USA}
\affiliation{Center for Exoplanets and Habitable Worlds, 525 Davey Laboratory, The Pennsylvania State University, University Park, PA, 16802, USA}

\author[0000-0001-9911-7388]{David W.\ Latham} 
\affiliation{Center for Astrophysics | Harvard \& Smithsonian, 60 Garden Street, Cambridge, MA 02138, USA}

\author[0000-0002-9082-6337]{Andrea S.J.\ Lin}
\affiliation{Department of Astronomy \& Astrophysics, 525 Davey Laboratory, The Pennsylvania State University, University Park, PA, 16802, USA}
\affiliation{Center for Exoplanets and Habitable Worlds, 525 Davey Laboratory, The Pennsylvania State University, University Park, PA, 16802, USA}

\author[0000-0002-1158-9354]{Gaspare Lo Curto} 
\affiliation{European Southern Observatory, Av.\ Alonso de Cordova 3107, Casilla 19001, Santiago de Chile, Chile}

\author[0000-0002-0651-4294]{Marcello Lodi} 
\affiliation{Fundaci\'on Galileo Galilei-INAF, Rambla Jos\'e Ana Fernandez P\'erez 7, 38712 Bre\~na Baja, Tenerife, Spain}

\author[0000-0002-9632-9382]{Sarah E.\ Logsdon}
\affiliation{NSF's National Optical-Infrared Astronomy Research Laboratory, 950 N.\ Cherry Ave., Tucson, AZ 85719, USA}

\author[0000-0001-7120-5837]{Christophe Lovis} 
\affiliation{Astronomy Department of the University of Geneva, 51 ch.\ des Maillettes, 1290 Versoix, Switzerland}

\author[0000-0001-9596-7983]{Suvrath Mahadevan}
\affiliation{Department of Astronomy \& Astrophysics, 525 Davey Laboratory, The Pennsylvania State University, University Park, PA, 16802, USA}
\affiliation{Center for Exoplanets and Habitable Worlds, 525 Davey Laboratory, The Pennsylvania State University, University Park, PA, 16802, USA}
\affiliation{ETH Zurich, Institute for Particle Physics \& Astrophysics, Zurich, Switzerland}   

\author[0000-0002-0048-2586]{Andrew Monson}
\affiliation{Steward Observatory, University of Arizona, 933 N.\ Cherry Ave, Tucson, AZ 85721, USA}

\author[0000-0001-8720-5612]{Joe P.\ Ninan}
\affil{Department of Astronomy and Astrophysics, Tata Institute of Fundamental Research, Homi Bhabha Road, Colaba, Mumbai 400005, India}

\author[0000-0002-9815-773X]{Francesco Pepe} 
\affiliation{Astronomy Department of the University of Geneva, 51 ch.\ des Maillettes, 1290 Versoix, Switzerland}

\author[0000-0002-9288-3482]{Rachael M.\ Roettenbacher}
\affiliation{Department of Astronomy, University of Michigan, 1085 S.\ University Ave., Ann Arbor, MI 48109, USA}  

\author[0000-0001-8127-5775]{Arpita Roy}
\affiliation{Space Telescope Science Institute, 3700 San Martin Dr, Baltimore, MD 21218, USA}
\affiliation{Department of Physics and Astronomy, Johns Hopkins University, 3400 N Charles St, Baltimore, MD 21218, USA}

\author[0000-0003-4422-2919]{Nuno C.\ Santos} 
\affiliation{Instituto de Astrof\'isica e Ci\^encias do Espa\c{c}o, Universidade do Porto, CAUP, Rua das Estrelas, 4150-762 Porto, Portugal}
\affiliation{Departamento de F\'isica e Astronomia, Faculdade de Ci\^encias, Universidade do Porto, Rua do Campo Alegre, 4169-007 Porto, Portugal}

\author[0000-0002-4046-987X]{Christian Schwab}
\affiliation{School of Mathematical and Physical Sciences, Macquarie University, Balaclava Road, North Ryde, NSW 2109, Australia}

\author[0000-0001-7409-5688]{Guðmundur Stef\'ansson}
\affiliation{Department of Astrophysical Sciences, Princeton University, 4 Ivy Lane, Princeton, NJ 08540, USA}
\affil{NASA Sagan Fellow} 

\author[0000-0002-4974-687X]{Andrew E.\ Szymkowiak}
\affiliation{Department of Astronomy, Yale University, 52 Hillhouse Ave., New Haven, CT 06511, USA}
\affiliation{Department of Physics, Yale University, 217 Prospect St, New Haven, CT 06511, USA}

\author[0000-0002-4788-8858]{Ryan C.\ Terrien}
\affiliation{Carleton College, One North College St., Northfield, MN 55057, USA}

\author[0000-0001-7576-6236]{Stephane Udry} 
\affiliation{Astronomy Department of the University of Geneva, 51 ch.\ des Maillettes, 1290 Versoix, Switzerland}

\author[0000-0002-5870-8488]{Sam A. Weiss}
\affiliation{Department of Astronomy, Yale University, 52 Hillhouse Ave., New Haven, CT 06511, USA}

\author[0000-0002-9216-4402]{François Wildi} 
\affiliation{Astronomy Department of the University of Geneva, 51 ch.\ des Maillettes, 1290 Versoix, Switzerland}

\author[0000-0002-0193-7132]{Thibault Wildi} 
\affiliation{Deutsches Elektronen-Synchrotron DESY, Notkestr. 85, 22607 Hamburg, Germany}

\author[0000-0001-6160-5888]{Jason T.\ Wright}
\affiliation{Department of Astronomy \& Astrophysics, 525 Davey Laboratory, The Pennsylvania State University, University Park, PA, 16802, USA}
\affiliation{Center for Exoplanets and Habitable Worlds, 525 Davey Laboratory, The Pennsylvania State University, University Park, PA, 16802, USA}
\affiliation{Penn State Extraterrestrial Intelligence Center, 525 Davey Laboratory, The Pennsylvania State University, University Park, PA, 16802, USA}